%% file: echo_chamber.tex
\documentclass[reprint,aps,pre,floatfix,twocolumn,showpacs,superscriptaddress
]{revtex4-1}
\pdfoutput=1
\usepackage{comment}
\usepackage{type1cm}     
\usepackage{graphicx}     
\usepackage{xspace}     
\usepackage{balance}     
\usepackage{booktabs}     
\usepackage{multi row}     
\usepackage{bold-extra}     
\usepackage{siunitx}          
\usepackage[vlined,linesnumbered,ruled,noend]{algorithm2e}     
\usepackage{microtype}    
\usepackage{units}     
\usepackage{mathtools}     
\usepackage{amssymb}     

 \usepackage{hyperref}
 \hypersetup{
    colorlinks=true,
    linkcolor=blue,
    citecolor=blue,
    filecolor=blue,
    urlcolor=blue,
}

\usepackage[table,xcdraw]{xcolor}
\usepackage{subcaption}
\usepackage[font=small,labelfont=bf,
tableposition=top,
   justification=justified,
   format=plain]{caption}
 
\newcommand{\spara}[1]{\smallskip\noindent\textbf{#1}}

\newcommand{\av}[1]{\langle #1 \rangle}

\begin{document}

\title{Echo Chambers on Social Media: A comparative analysis}

\author{Matteo Cinelli}
\email{matteo.cinelli@uniroma2.it}
\affiliation{Applico Lab -- ISC CNR, Via dei Taurini 19, 00185 Roma, Italy}
\author{Gianmarco De Francisci Morales}
\email{gdfm@acm.org}
\affiliation{ISI Foundation, via Chisola 5, 10126 Torino, Italy}
\author{Alessandro Galeazzi}
\email{a.galeazzi002@unibs.it}
\affiliation{University of Brescia, Via Branze, 59, 25123 Brescia , Italy}
\author{Walter Quattrociocchi}
\email{w.quattrociocchi@unive.it}
\affiliation{Ca'Foscari Univerity of Venice, via Torino 155, 30172 Venezia, Italy}
\author{Michele Starnini}
\email{michele.starnini@gmail.com}
\affiliation{ISI Foundation, via Chisola 5, 10126 Torino, Italy}

\begin{abstract}
Recent studies have shown that online users tend to select information adhering to their system of beliefs, ignore information that does not, and join groups -- i.e., echo chambers -- around a shared narrative.  
Although a quantitative methodology for their identification is still missing, the phenomenon of echo chambers is widely debated both at scientific and political level.
To shed light on this issue, we introduce an operational definition of echo chambers and perform a massive comparative analysis on more than 1B pieces of contents produced by 1M users on four social media platforms: Facebook, Twitter, Reddit, and Gab. 
We infer the leaning of users about controversial topics -- ranging from vaccines to abortion -- and reconstruct their interaction networks by analyzing different features, such as shared links domain, followed pages, follower relationship and commented posts.  
Our method quantifies the existence of echo-chambers along two main dimensions: homophily in the interaction networks and bias in the information diffusion toward likely-minded peers.
We find peculiar differences across social media.
Indeed, while Facebook and Twitter present clear-cut echo chambers in all the observed dataset, Reddit and Gab do not.
Finally, we test the role of the social media platform on news consumption by comparing Reddit and Facebook.
Again, we find support for the hypothesis that platforms implementing news feed algorithms like Facebook may elicit the emergence of echo-chambers.

\end{abstract}

\maketitle

\section*{Introduction}
\label{sec:intro}

Social media allow users to access and share an unprecedented amount of information, thus changing the way we interact, debate and form our opinions~\cite{quattrociocchi2017part, nguyen2019testing}.
The wide availability of content sparked the enthusiastic idea that users might be better informed and exposed to diversified point of views \cite{dubois2018echo,bode2016political,newman2019reuters}.
However, the human attention span remains limited~\cite{baronchelli2018emergence, cinelli2020selective} and news feed algorithms might influence the selection process by promoting content similar to the ones already seen, thus reducing content diversity~\cite{schmidt2017anatomy,cinelli2020covid} and eventually leading to polarization~\cite{conover2011political,bail2018exposure,perra2019modelling,sasahara2019inevitability}.
On top of opinion polarization, users show the tendency to select information that adheres to their beliefs and join polarized groups formed around a shared narrative, called echo chambers~\cite{del2016spreading,jamieson2008echo,garrett2009echo,quattrociocchi2016echo,garimella2018political}.
Inside these closed communities formed by users having similar preferences and content consumption patterns, the information spreading is often biased~\cite{bessi2015science,del2016spreading,garimella2017effect,cota2019quantifying,balsamo2019inside,cossard2020falling}. 
Such a configuration might hamper the democratic deliberative process by altering the way facts are perceived \citep{sunstein2009republic}.

Nowadays, echo chambers are one of the most debated issues in relation to the social media environment~\cite{jamieson2008echo,quattrociocchi2017inside}, given their potential role in fostering actions of political manipulation and influence on voting behavior~\cite{shao2017spread,lazer2018science,bovet2019influence}.
Some studies point out echo chambers as an emerging effect of human tendencies, such as selective exposure and contagion~\cite{himelboim2013birds,flaxman2016filter,nikolov2015measuring}.
Moreover, group polarization theory~\citep{sunstein2002law} has been proposed as a mechanism to model the dynamics leaning to the emergence of echo-chambers in online social networks~\cite{baumann19}. 
It is remarkable that major social media and former U.S. Presidents alike have voiced such concerns \footnote{E.g., Obama foundation's attempt to address the issue of echo chambers.~\url{https://www.engadget.com/2017/07/05/obama-foundation-social-media-echo-chambers} \\ Facebook's CEO Mark Zuckerberg's open letter.~\url{https://www.facebook.com/notes/mark-zuckerberg/building-global-community/10103508221158471/}}.
Recently, however, the effects and the very existence of echo chambers have been questioned~\cite{barbera2015tweeting,dubois2018echo,bruns2017echo,bruns2019filter}.
This heated debate calls for a quantitative analysis able to gauge the presence of echo chambers across topics and social media.  
In this work, we provide a formal assessment of echo chambers by introducing an operational definition independent of the social media platform considered.

We propose a definition of echo chambers based on the coexistence of two main ingredients:
($i$) opinion polarization with respect to a controversial topic, and 
($ii$) homophilic interactions between users, i.e., the preference to interact with like-minded peers.
We operationalize these two abstract concepts into observables that can be quantified empirically, thus providing a common methodological ground to obtain consistent results and compare them across different social media.
We perform a comparative analysis on four major social media platforms: Facebook, Twitter, Reddit, and Gab. 
These media share some common features and functionalities (e.g., they all allow social feedback actions such as likes or upvotes) and design (e.g., Gab is similar to Twitter) but also distinctive features (e.g., Reddit is organized in communities of interest called subreddits).

While the environment and the main features behind mainstream social media have been widely investigated, other social media such as Reddit and especially Gab have been somewhat overlooked.
Reddit is one of the most visited websites worldwide \footnote{\url{https://www.alexa.com/siteinfo/reddit.com}} and is organized as a forum to collect discussions about a wide amount of topics, ranging from politics to emotional support.
Gab is relatively different, as it claims to be a social platform aimed at protecting free speech.
Such a claim, together with the political leaning of its developers, made Gab the ``safe haven'' for the alt-right movement.
However, low moderation and regulation on content has resulted in widespread hate speech.
For these reasons, it has been repeatedly suspended by its service provider, and its mobile app was banned from both App and Play stores~\cite{zannettou2018gab}.
All these features make the comparison of the aforementioned social media particularly interesting for investigation.
Overall, we take into account the interactions of more than 1M active users on the four platforms, for a total of more than 1B unique pieces of content, including posts and social interactions.
Our findings suggest that platforms organized around social networks and with algorithms accounting for social feedback may increase polarization and favor the emergence of echo chambers.

\section*{Characterizing echo chambers in social media}

At an abstract level, the echo chamber phenomenon can be understood in the context of selective exposure theory~\citep{klapper1960effects}.
Humans have a tendency to seek information adhering to their pre-existing opinion, a phenomenon sometimes referred to as \emph{confirmation bias} \cite{nickerson1998confirmation}. Such a tendency has been proven to be dominant in content consumption on online social media \cite{del2016echo,bessi2015science,del2016spreading,garimella2017effect}.
In a social context, this tendency may foster the emergence of homophilic clusters of individuals.
This, in turn, creates an environment where individuals are surrounded by people whose opinion agrees with their own: an \emph{echo chamber}.

A fundamental mechanism to explain the origin of the tendency to selective exposure can be found in cognitive dissonance theory~\citep{festinger1962theory}.
The theory posits that individuals strive towards internal consistency of thoughts and beliefs, by virtue of the fact that inconsistency, or \emph{dissonance}, is psychologically uncomfortable.
An individual will thus try to avoid information and situations that are likely to increase their dissonance, and seek instead consonant ones.
Cognitive dissonance is thus possibly the \emph{primum movens}, or innate root cause, of the ultimate formation of echo chambers.
According to group polarization theory~\citep{sunstein2002law}, an echo chamber can act as a mechanism to reinforce an existing opinion within a group, and as a result move the entire group towards more extreme opinions.
The lack of exposure to alternative opinions also creates a false perception of unanimity, and thus a different perception of reality across groups, which may hinder the democratic debate given the lack of a shared common ground on which to operate.

\subsection*{Operational definitions}

An echo-chamber can be defined as an environment in which the opinion, political leaning, or belief of an individual about a certain topic are reinforced due to repeated interactions with peers who share similar points of view.
Two key elements are needed for this scenario to take place.
First, a group of individuals that share a common opinion in opposition to other individuals or groups characterized by different attitudes regarding the same topic. 
Second, social interactions that convey a flow of information between these individuals about the topic under consideration, that can thus influence their beliefs on the subject. 
 Such interactions are more likely to be established between individuals characterized by similar opinions, that is, there is a certain degree of homophily in social interactions. 
Therefore, echo-chambers are characterized by the coexistence of two elements: ($i$) opinion \emph{polarization} with respect to  a controversial topic, and ($ii$) \emph{homophily} in interactions, i.e. the preference to interact with like-minded peers. 
These two abstract concepts need to be operationalized in order to be gauged on empirical social systems, and in particular within the specific context of online social media. 

In order to quantify the degree of polarization in social systems, one needs first to identify the attitude of users, at a micro level. 
On online social media, the \emph{individual leaning} of a user $i$ toward a specific topic, $x_i$, can be inferred in different ways, via the content produced, or the endorsement network among users~\citep{garimella2016quantifying,garimella2018quantifying}.
With respect to the content, its leaning can be defined as the attitude expressed by a piece of content towards a specific topic. 
This leaning can be explicit (e.g., arguments supporting a narrative) or implicit (e.g., framing and agenda setting).
Let us consider a user $i$ producing a number $a_i$ of contents, $\mathcal{C}_i = \{ c_1, c_2, \ldots, c_{a_i} \}$, where $a_i$ is the \emph{activity} of user $i$ and each content leaning is assigned a numeric value.
Then the individual leaning of user $i$ can be defined as the average of the leanings of contents produced,
\begin{equation}
  x_i \equiv \frac{\sum_{j=1}^{a_i} c_j}{a_i}.
  \label{eq:ind_lean}
\end{equation}

Once individual leanings are inferred, polarization can be defined as a state of the system such that the distribution of leanings, $P(x)$, is heterogeneous.
If opinions are assumed to be embedded in a one-dimensional space, as usual in case of topics characterized by positive versus negative stances, polarization can be quantified by a bimodal distribution.
That is, if opinions are represented on an axis, $x \in [-1, +1]$, without loss of generality, polarization is then characterized by two well-separated peaks in $P(x)$, for positive and negative opinions, while neutral ones are absent or underrepresented in the population. 
Note that polarization can happen independently from the structure or the very presence of social interactions.

Homophily in social interactions can be quantified by representing interactions as a social network, and then analyzing its structure with respect to the opinions of the agents \cite{kossinets2009origins,aiello2012friendship,bessi2016homophily,cota2019quantifying}.  
From online social media, social networks can be reconstructed in different ways, where links represent social relationships or interactions. 
Since we are interested in capturing the possible exchange of opinion between users, we assume directed links to represent the substrate over which information may flow. 
For instance, if user $i$ follows user $j$ on Twitter, user $i$ can see tweets produced by user $j$, thus there is a flow of information from node $j$ to node $i$ in the network. 
That is, when the reconstructed network is directed, we assume the link direction points to possible influencers (opposite of information flow).
Actions such as mentions or retweets may convey similar flows.  
In some cases, direct relations between users are not available in the data, so one needs to assume some proxy for social connections, e.g., a link between two users if they comment the same post on Facebook. 
Crucially, the two elements characterizing the presence of echo-chambers, polarization and homophilic interactions, should be quantified independently.

\subsection*{Implementation on social media}

This section explains how we implement the operational definitions defined above on different social media. 
For each medium, we detail $(i)$ how we quantify the individual leaning of users, and $(ii)$ how we reconstruct the interaction network on top of which the information spread. Further details are provided in the Materials and Methods Section.

\spara{Twitter}.  We consider the set of tweets posted by user $i$ that contain links to news organizations of known political leaning. 
To each news organization is associated a political leaning score \cite{Bakshy1130} ranging from extreme left to extreme right in accordance to the classification reported in Materials and Methods.
We infer the individual leaning of a user $i$, $x_i \in [-1,+1]$ by averaging the scores of the news organizations linked by user $i$ according to Eq. \eqref{eq:ind_lean}.
We analyze three different data sets collected on Twitter related to controversial topics: gun control, Obamacare, and abortion. 
For each data set, the social interaction network is reconstructed by using the following relation, so that there exists a direct link from node $i$  to node $j$ if user $i$ follows user $j$. 
Henceforth we focus on the data set about abortion, others are shown in the Supplementary Material (SM).

\spara{Facebook}. 
The individual leaning of users is quantified by considering endorsements in the form of likes to posts.
While other actions such as comments or shares could be taken into account, the written text may radically change the inferred leaning.
Additionally, while a like is usually a positive feedback on a news item, comments and share can be associated to different purposes~\cite{schmidt2017anatomy}.
A comment can have multiple features and meanings and can generate collective debate, while a share indicates a desire to spread a news item to friends.  
Posts are produced by pages that are labeled in a certain number of categories, and to each category we assign a numerical value (e.g., Anti-Vax (+1) or Pro-Vax (-1)).
Each like to a post (only one like per post is allowed) represents an endorsement for that content, which is assumed to be aligned with the labeling of the page. 
Thus, the individual leaning of a user is defined as the average of the content leanings of the posts liked by the user, according to Eq. \eqref{eq:ind_lean}.

We analyze three different data sets collected on Facebook regarding a specific topic of discussion: {vaccines}, science versus conspiracy, and news. 
The interaction network is defined by considering comments.
In such an interaction network two users are connected if they co-commented at least one post.
Henceforth we focus on the data set about vaccines and news, others are shown in the SM.


\spara{Reddit}. 
Here, the individual leaning of users is quantified similarly to Twitter, by considering the links to news organizations in the content produced by the users, submissions and comments. 
The interaction network is defined by considering comments and submissions, by reconstructing the information flow.
There exists a direct link from node $i$ to node $j$ if user $i$ comments on a submission or comment by user $j$ (we assume that $i$ reads the comment they are replying to, which is written by $j$). 
We analyze three data sets collected on different subreddits: the\_donald, politics, and news. In the following we focus on the data set collected on the Politics and on the News subreddit, others are shown in the SM.


\spara{Gab}. 
The political leaning $x_i$ of user $i$ is computed by considering the set of contents posted by user $i$ that contain a link to news organizations of known political leaning, similarly to Twitter and Reddit. 
To obtain the leaning $x_i$ of user $i$, we averaged the scores of each link posted by user $i$ according to Eq. \eqref{eq:ind_lean}.
The interaction network is  reconstructed by exploiting the co-commenting relationships under posts in the same way as for Facebook. 
Given two users $i$ and $j$, an undirected edge between $i$ and $j$ exists if and only if they comment under the same post.

%

\section*{Comparative Analysis}
\label{sec:res}

In the following we compare the presence or absence of echo-chambers across social media. 
We select one data set for each social media: Abortion (Twitter), Vaccines (Facebook), Politics (Reddit), and Gab as a whole. Results for other data sets for the same medium are qualitatively similar, as shown in the SM.
We first characterize echo-chambers in the topology of the networks, then look at their effects on information diffusion. 
Finally, we directly compare Facebook and Reddit on a common topic, news consumption, to highlight the differences in the behavior of users.

 \begin{figure}[tbp]
        \centering
        \begin{subfigure}[t]{0.23\textwidth}   
            \centering 
            \includegraphics[width=\textwidth]{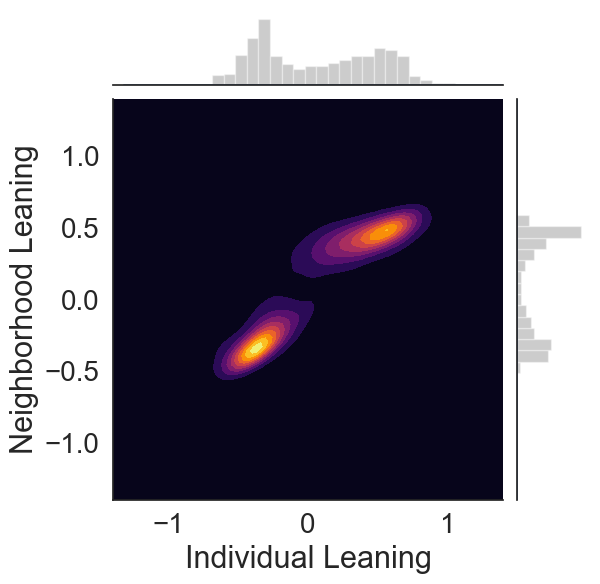}
            \caption{Twitter}    
            \label{fig:abortion_nn}
        \end{subfigure}
        \hfill
        \begin{subfigure}[t]{0.23\textwidth}   
            \centering 
            \includegraphics[width=\textwidth]{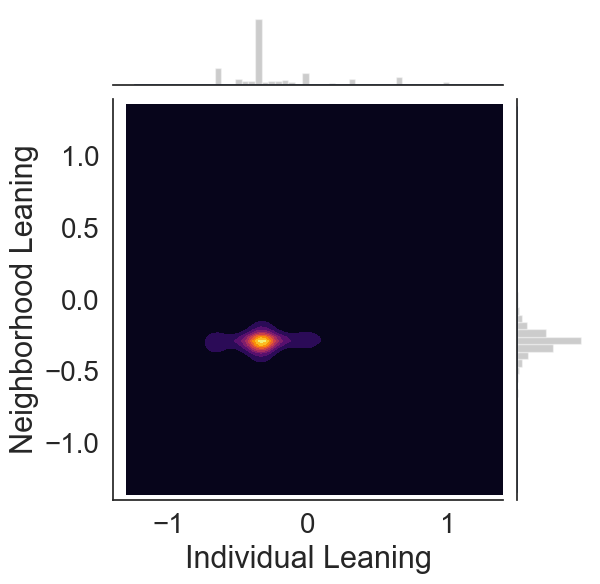}
            \caption{Reddit}    
            \label{fig:politics_nn}
        \end{subfigure}

        \begin{subfigure}[t]{0.23\textwidth}
            \centering
            \includegraphics[width=\textwidth]{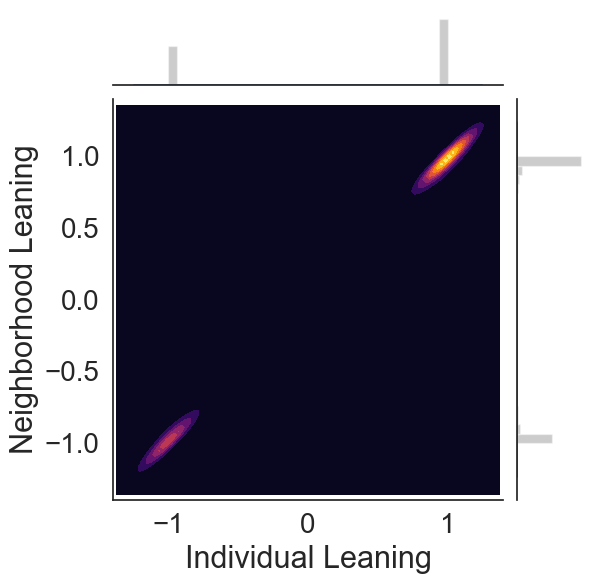}
            \caption{Facebook}
            \label{fig:vax_nn}
        \end{subfigure}
        \hfill
        \begin{subfigure}[t]{0.23\textwidth}  
            \centering 
            \includegraphics[width=\textwidth]{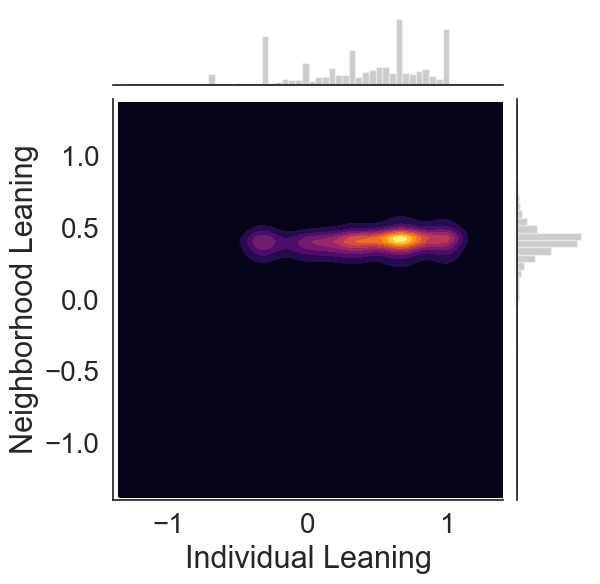}
            \caption{Gab}    
            \label{fig:gab_nn}
        \end{subfigure}
        \caption{Joint distribution of the leaning of users $x$ and the average leaning of their neighborhood $x^{NN}$ for different data sets.   Colors represent the density of users: the lighter, the larger the number of users.  Marginal distribution $P(x)$ and $P^{N}(x)$ are plotted on the x and y axis, respectively. }
        \label{fig:pol_in_out}
    \end{figure}

\subsection*{Homophily in the interaction networks}

The topology of the interaction network can reveal the presence of echo-chambers, where users are surrounded by peers with similar leaning and thus are exposed with higher probability to similar contents. 
In network terms, this translates into a node $i$ with a given leaning $x_i$ more likely to be connected with nodes with a leaning close to $x_i$ \cite{cota2019quantifying}. 
This concept can be quantified by defining, for each user $i$, the average leaning of their neighborhood, as $x_i^{N} \equiv \frac{1}{k_{i}^{\rightarrow}} \sum_j A_{ij} x_j$, where $A_{ij}$ is the adjacency matrix of the interaction network, $A_{ij} = 1$ if there is a link from node $i$ to node $j$, $A_{ij} = 0$ otherwise, and  $k_{i}^{\rightarrow} = \sum_jA_{ij}$ is the out-degree of node $i$.

Fig.~\ref{fig:pol_in_out} shows the correlation between the leaning of a user $i$ and the leaning of their neighbors, $x_i^{N}$, for the four social media under consideration.
The probability distributions $P(x)$ (individual leaning) and $P^{N}(x)$ (average leaning of neighbors) are plotted on the x and y axis, respectively.
All plots are color-coded contour maps, which represent the number of users in the phase space $(x, x^{N})$: the brighter the area in the map, the larger the density of users in that area.
The topics of vaccines and abortion, on Facebook and Twitter, respectively, 
clearly show two distinct groups whose leanings differ quite starkly, as indicated by the two bright areas characterized by a high density of users with like-minded neighbors.
Similar behavior is found for different topics from the same social media platform, see SM.
Conversely, Reddit and Gab show a different picture. 
The corresponding plots in Fig.~\ref{fig:pol_in_out} display a single bright area, indicating that users do not split into groups with opposite leaning but form a single community, biased to the left (Reddit) or the right (Gab).
Similar results are found for different data sets on Reddit, see SM.

\begin{figure}[tbp]
        \centering
        \begin{subfigure}[t]{0.23\textwidth}   
            \centering 
            \includegraphics[width=\textwidth]{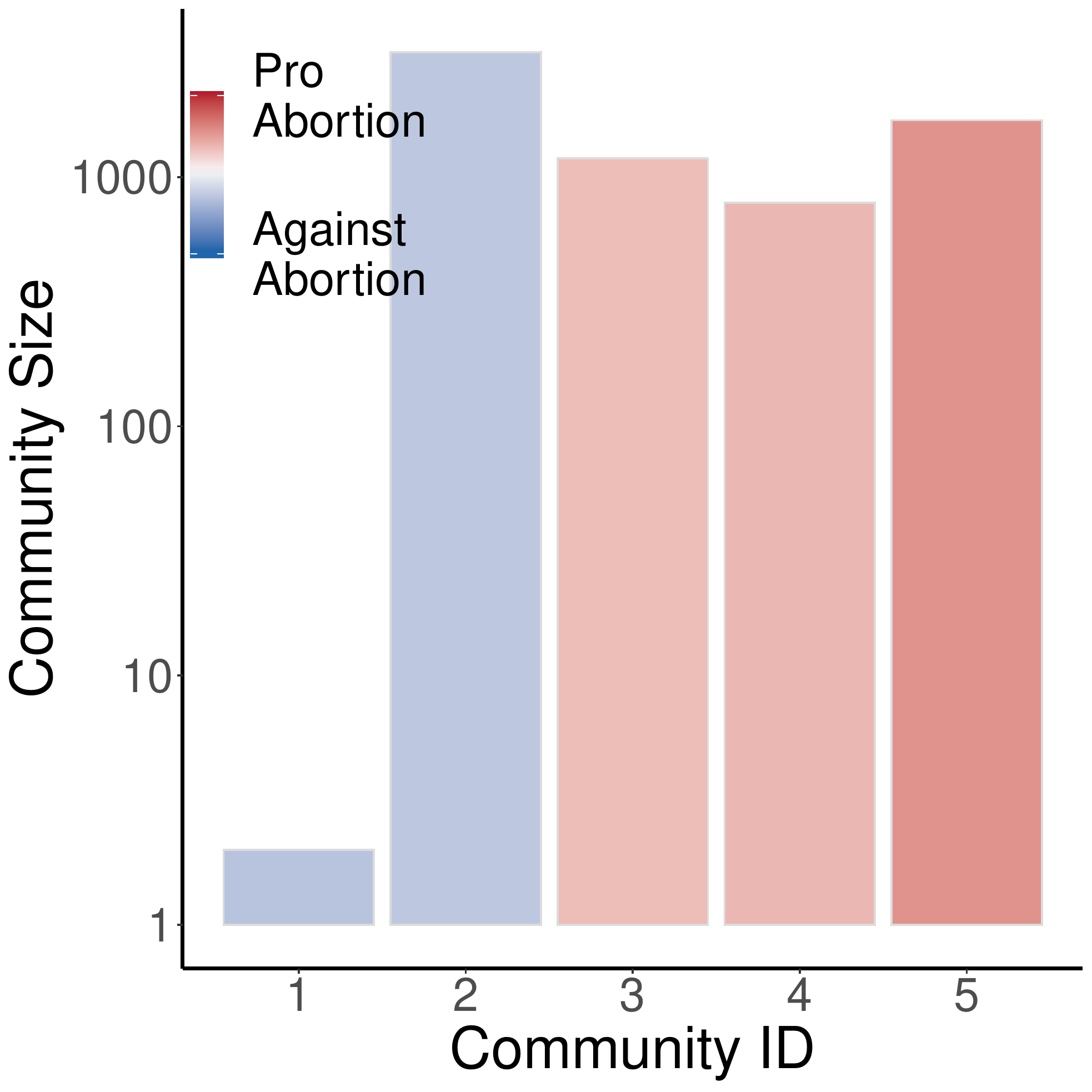}
            \caption{Twitter}
            \label{fig:abortion_comm}
        \end{subfigure}
        \hfill
        \begin{subfigure}[t]{0.23\textwidth}   
            \centering 
            \includegraphics[width=\textwidth]{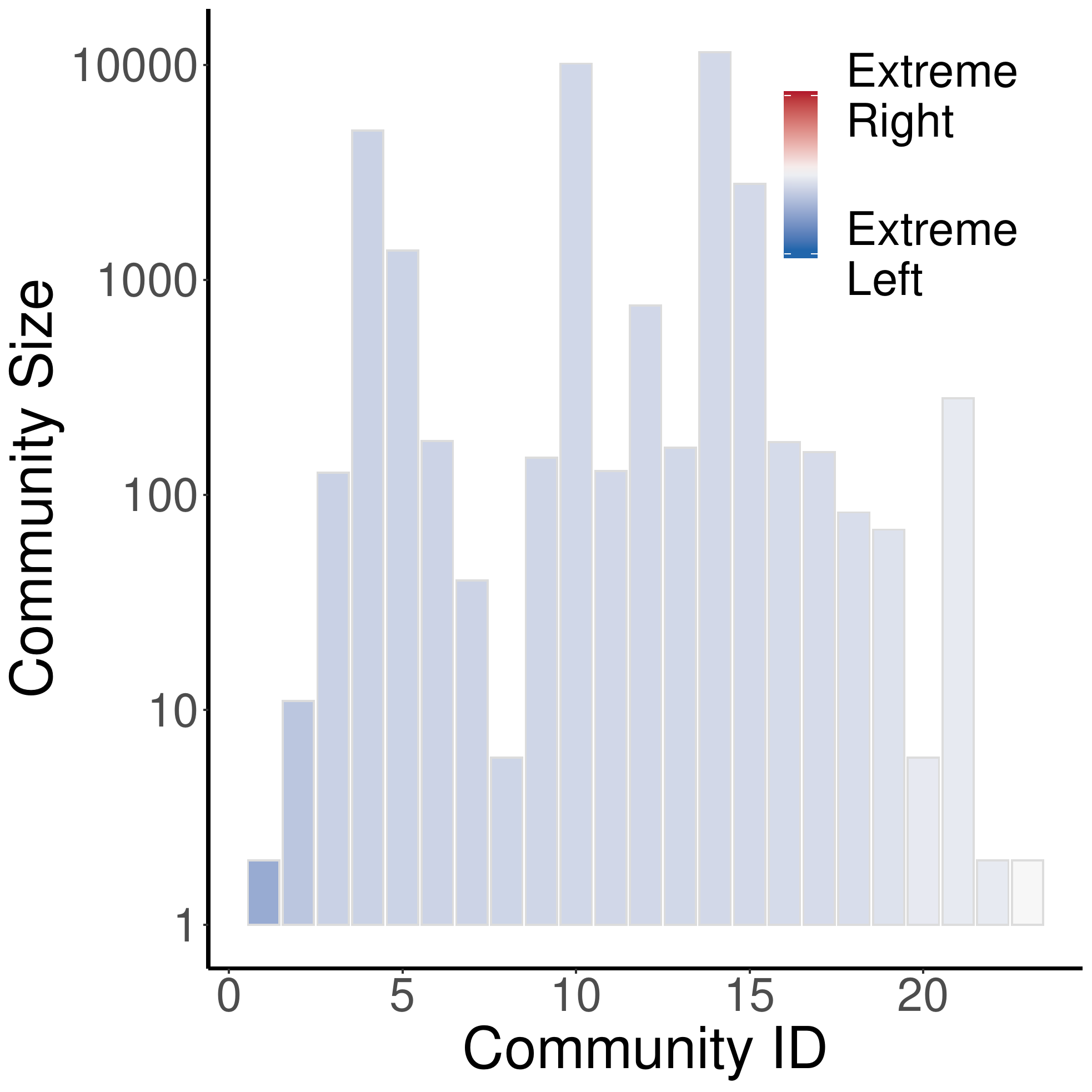}
            \caption{Reddit}
            \label{fig:politics_comm}
        \end{subfigure}

        \begin{subfigure}[t]{0.23\textwidth}
            \centering
            \includegraphics[width=\textwidth]{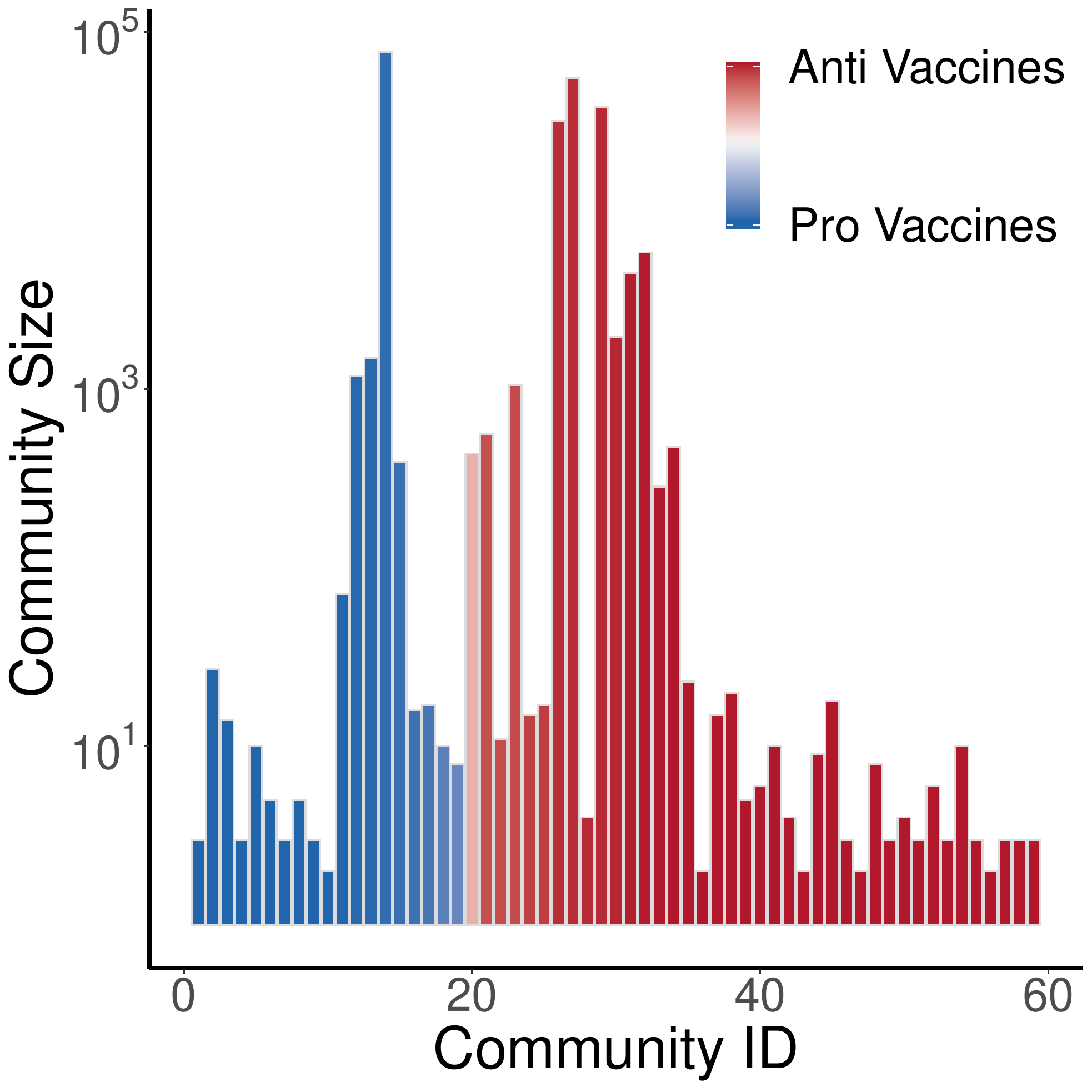}
            \caption{Facebook}
            \label{fig:vax_comm}
        \end{subfigure}
        \hfill
        \begin{subfigure}[t]{0.23\textwidth}  
            \centering 
            \includegraphics[width=\textwidth]{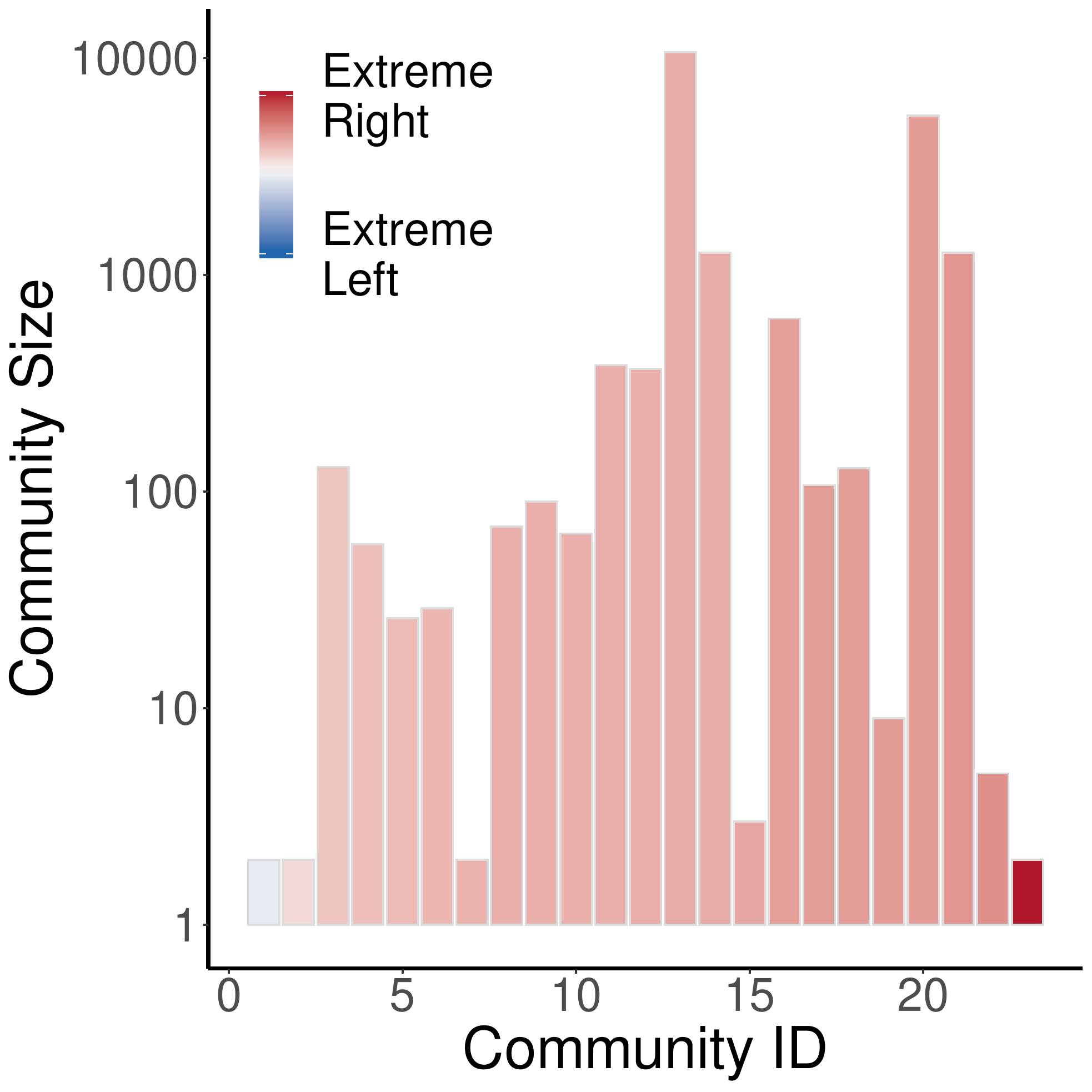}
            \caption{Gab}
            \label{fig:gab_comm}
        \end{subfigure}
        \caption{Size and average leaning of communities detected in different data sets. } 
        \label{fig:communities}
    \end{figure}


Homophilic interactions can be revealed by the community structure of the interaction networks. 
We detect communities by applying the Louvain algorithm for community detection~\citep{blondel2008fast}.
We remove singleton communities with only one user and look at the average leaning of each community, determined as the average of individual leanings of its members.

Fig.~\ref{fig:communities} shows the communities emerging for each social medium, arranged by increasing average leaning on the x-axis (color-coded from blue to red), while the y-axis reports the size of the community.
We find a picture that confirms the pattern observed before. 
On Facebook and Twitter, communities span the whole spectrum of possible leanings, but each community is formed by users with similar leaning. 
Some communities are characterized by very strong average leaning, especially in the case of Facebook.
Conversely, communities on Reddit and Gab do not cover the whole spectrum, and all show similar average leaning. 
Furthermore, it is noticeable the almost total absence of communities with leaning very close to 0, confirming the polarized state of the systems. 
In addition, the number of communities identified is different among the four social media. 
The similar number of communities found in Gab and Reddit and the strong difference with respect to Facebook and Twitter suggests that both platforms structure and feedback algorithm may have an impact on the clustering process of users.


\subsection*{Effects on information spreading}
\label{sec:spreadability}

The presence of echo chambers can be gauged by simple models of information spreading:
users are expected to exchange information more likely with peers sharing a similar leaning \cite{cota2019quantifying, garimella2017reducing, garimella2016quantifying}.
Classical epidemic models such as the susceptible-infected-recovered (SIR)
model~\cite{anderson92} have been used to study the diffusion of information, such as rumors or news~\cite{ZHAO2013995,PhysRevLett.111.128701}.
In the SIR model, each agent can be in either of three states: susceptible (unaware of the circulating information), infectious (aware and willing to spread it further), or recovered (aware but not willing to transmit it anymore). 
Susceptible (unaware) users may become infectious (aware) upon contact with infected neighbors, with certain transmission probability $\beta$.
Infectious users can spontaneously become recovered with probability $\nu$.
In order to measure the effects of the leaning of users on the diffusion of information, 
we run the SIR dynamics on the interaction networks, 
by starting the epidemic process with only one node $i$ infected, and stopping it when no more infectious nodes are left.

The set of nodes in a recovered state at the end of the dynamics started with user $i$ as seed of infection, i.e., those that become aware of the information initially propagated by user $i$, forms the \textit{set of influence} of user $i$, $\mathcal{I}_i$~\cite{PhysRevE.71.046119}. 
The set of influence of a user thus represents those individuals that can be reached by a piece of content sent by him{/her}, depending on the effective infection ratio $\beta/\nu$.
One can compute the average leaning of the set of influence of user $i$, $\mu_i$, as
\begin{equation}
  \mu_{i}\equiv |\mathcal{I}_i|^{-1}  \sum_{j \in \mathcal{I}_i} { x_j}.
\end{equation}
The quantity $\mu_i$ indicates how polarized are the users that can be reached by a message initially propagated by user $i$ \cite{cota2019quantifying}.

\begin{figure}[tbp]
        \centering
        \begin{subfigure}[t]{0.23\textwidth}   
            \centering 
            \includegraphics[width=\textwidth]{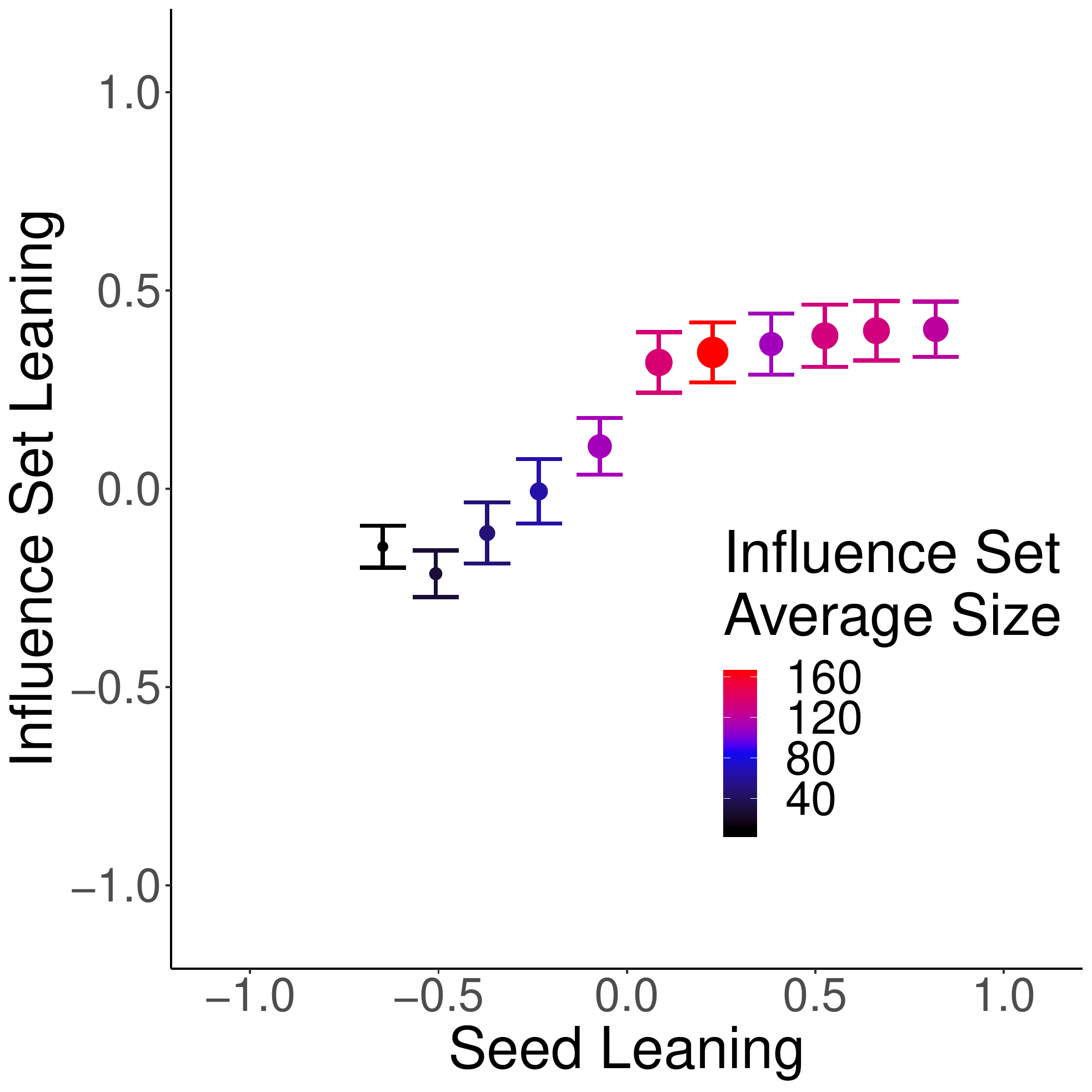}
            \caption{Twitter}    
            \label{fig:abortion_SIR}
        \end{subfigure}
        \hfill
        \begin{subfigure}[t]{0.23\textwidth}   
            \centering 
            \includegraphics[width=\textwidth]{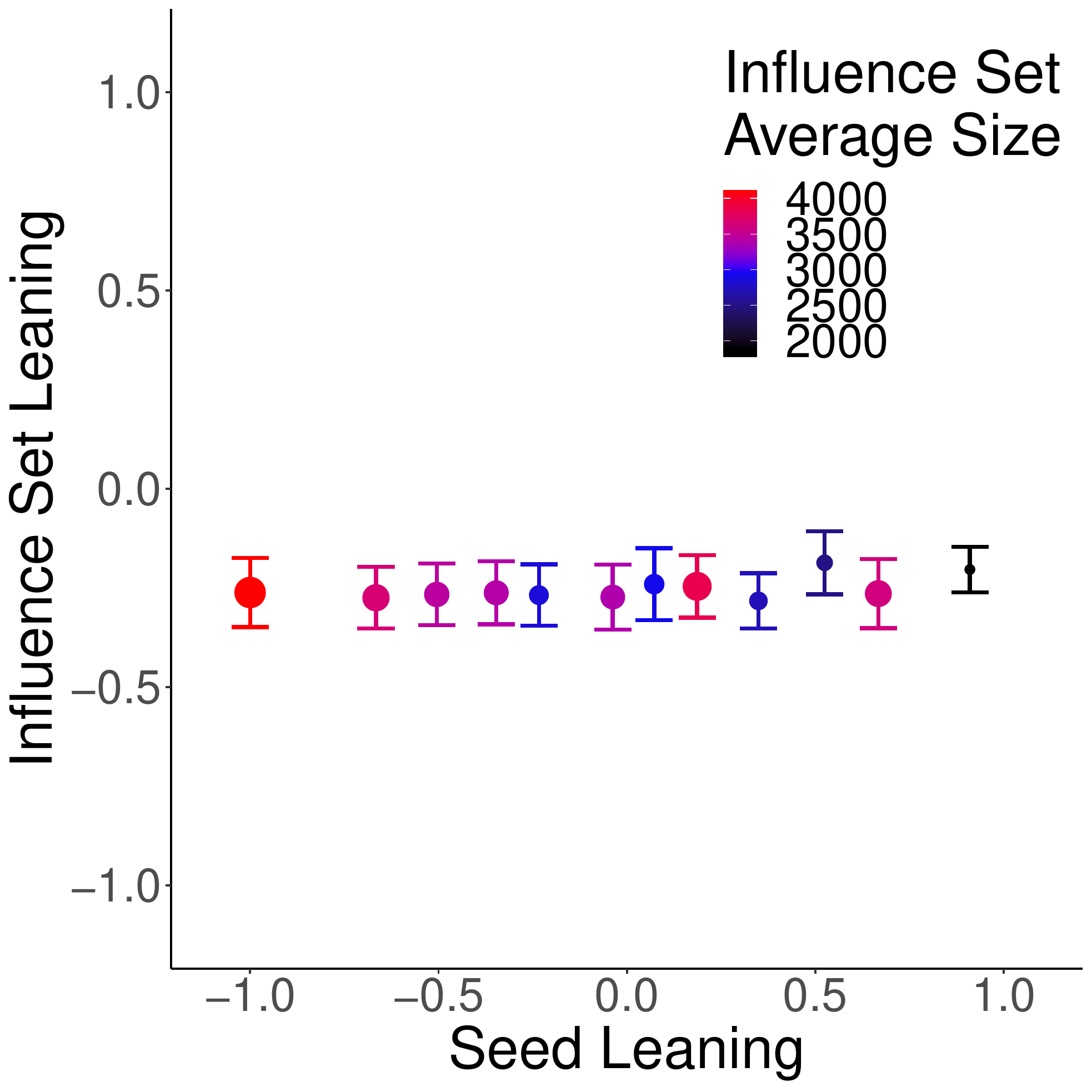}
            \caption{Reddit}    
            \label{fig:politics_SIR}
        \end{subfigure}

        \begin{subfigure}[t]{0.23\textwidth}
            \centering
            \includegraphics[width=\textwidth]{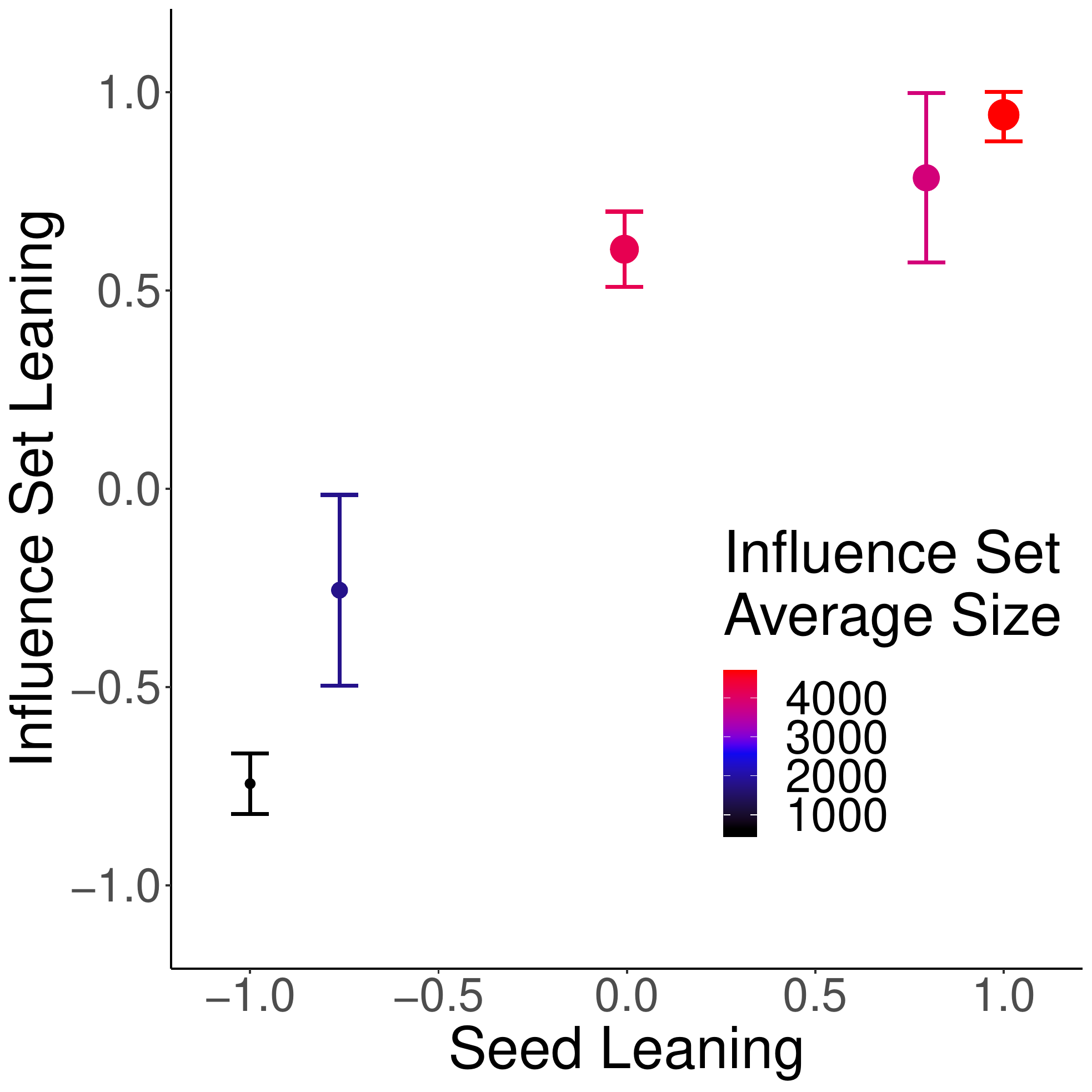}
            \caption{Facebook}
            \label{fig:vax_SIR}
        \end{subfigure}
        \hfill
        \begin{subfigure}[t]{0.23\textwidth}  
            \centering 
            \includegraphics[width=\textwidth]{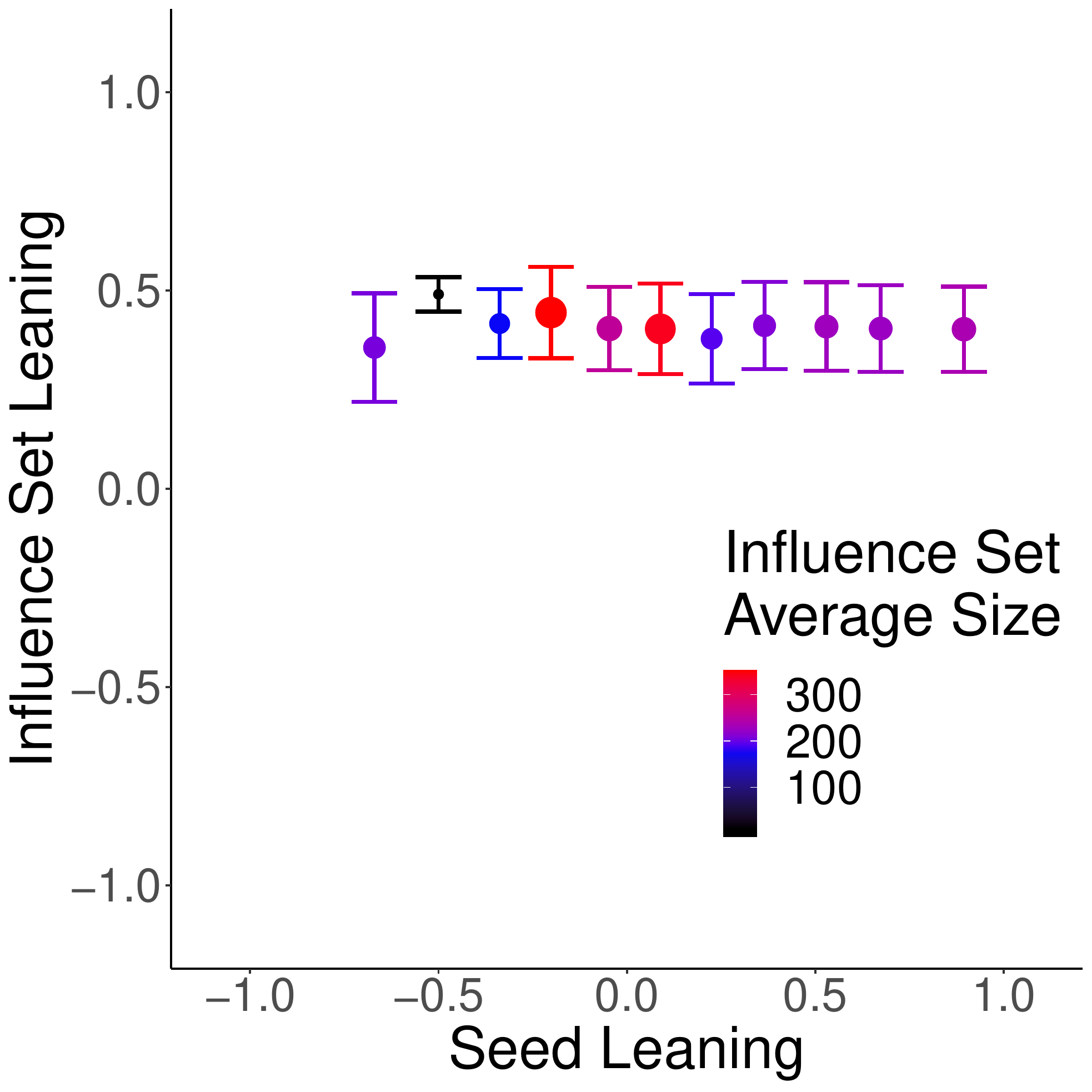}
            \caption{Gab}    
            \label{fig:gab_SIR}
        \end{subfigure}
        \caption{ Average leaning $\av{\mu(x)}$ of the influence sets reached by users with leaning $x$, for different data sets under consideration. 
        Size and color of each point represents the average size of the influence sets. The parameters of the SIR dynamics are set to $\beta=0.10\av{k}^{-1}$ for panel (a), $\beta=0.01\av{k}^{-1}$ for panel (b), $\beta=0.05\av{k}^{-1}$ for panel (c) and  $\beta=0.05\av{k}^{-1}$ for panel (d), while $\nu$ is fixed at 0.2 for all simulations.}
        \label{fig:sir}
    \end{figure}

Fig.~\ref{fig:sir} shows the average leaning $\av{\mu(x)}$ of the influence sets reached by users with leaning $x$, for the different data sets under consideration.
The recovery rate $\nu$ is fixed at 0.2 for every dataset, while relationship between infection rate $\beta$ and average degree $\av{k}$ vary from dataset to dataset and is reported in the caption of each figure. 
More details about the network used for the SIR model are reported in Table~\ref{tab:data} in Methods and Material Section.
Again, one can observe a clear distinction between Facebook and Twitter, on one side, and Reddit and Gab on the other side. 
For the topics of vaccines and abortion, on Facebook and Twitter, respectively, users with a given leaning are much more likely to be reached by information propagated by users with similar leaning, i.e., $\av{\mu(x)} \sim x$. Similar behavior is found for different topics from the same social media platform, see SM. 
Conversely, Reddit and Gab show a different behavior: the average leaning of the set of influence, $\av{\mu(x)}$, does not depend on the leaning $x$. 
These results indicate that in some social media, namely Twitter and Facebook,  information diffusion is biased toward individuals that share similar leaning, while in others -- Reddit and Gab in our analysis --  this effect is absent. 
The quantity $\av{\mu(x)}$, indeed, gauges the strength of the echo chambers effect: the more $\av{\mu(x)}$ is
close to $x$, the stronger the echo chamber effect, while if $\av{\mu(x)}$ is independent of $x$, echo-chambers are not present. 
Our results are robust with respect to different values of the effective infection ratio $\beta/\nu$, see SM.

Furthermore, Fig. \ref{fig:sir} shows that the spreading capacity, represented by the average size of the influence sets (color coded in Fig. \ref{fig:sir}), depends on the leaning of the users.
On Twitter, pro-abortion users are more likely to reach larger audiences, the same is true for anti-vax users on Facebook, left-leaning users on Reddit, and right-leaning users on Gab (in this data set, left-leaning users are almost absent though).

    \subsection*{News Consumption on Facebook and Reddit}

The striking differences observed across social media, in terms of homophily in the interaction networks and information diffusion, could be attributed to different topics taken into account. 
For this reason, here we compare Facebook and Reddit on a common topic, news consumption. 
Facebook and Reddit are particularly apt to a cross-comparison since they share the definition of individual leaning (computed by using the classification provided by mediabiasfactcheck.org, see Methods for further details) and the rationale in creating connections among users that is based on an interaction network.

\begin{figure}[tbp]
        \centering
        \begin{subfigure}[t]{0.23\textwidth}
            \centering 
            \includegraphics[width=\textwidth]{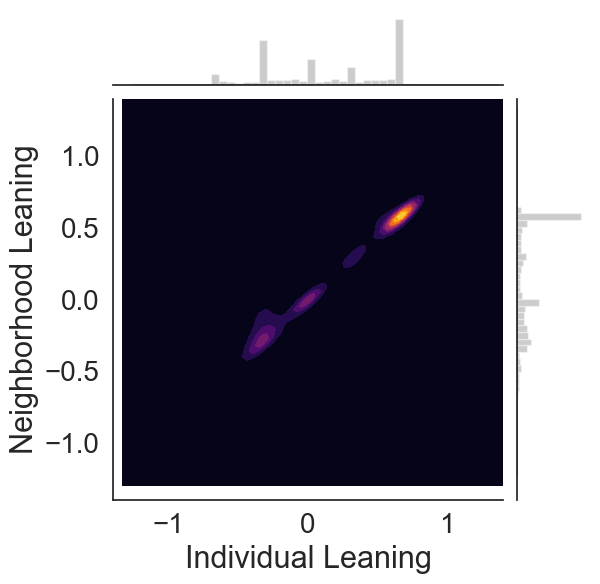}
            \label{fig:fb_news_nn}
        \end{subfigure}
        \hfill
        \begin{subfigure}[t]{0.23\textwidth}   
            \centering 
            \includegraphics[width=\textwidth]{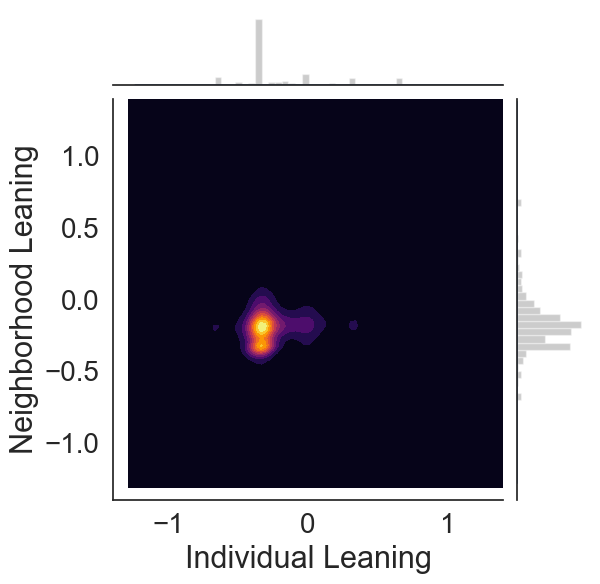}
            \label{fig:reddit_news_nn}
        \end{subfigure}
        \centering
        \begin{subfigure}[t]{0.23\textwidth}
            \centering 
            \includegraphics[width=\textwidth]{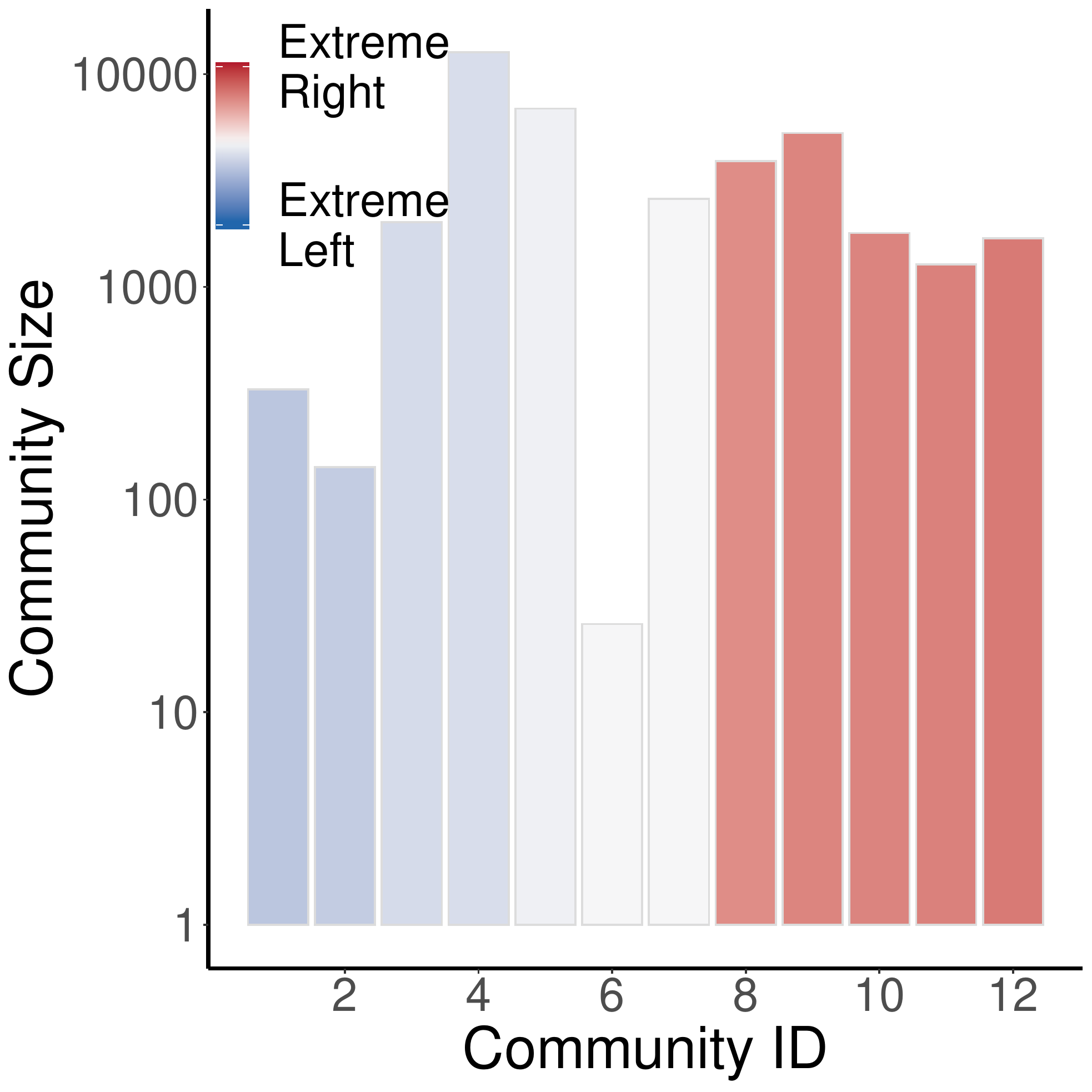}
            \label{fig:fb_news_SIR}
        \end{subfigure}
        \hfill
        \begin{subfigure}[t]{0.23\textwidth}   
            \centering 
            \includegraphics[width=\textwidth]{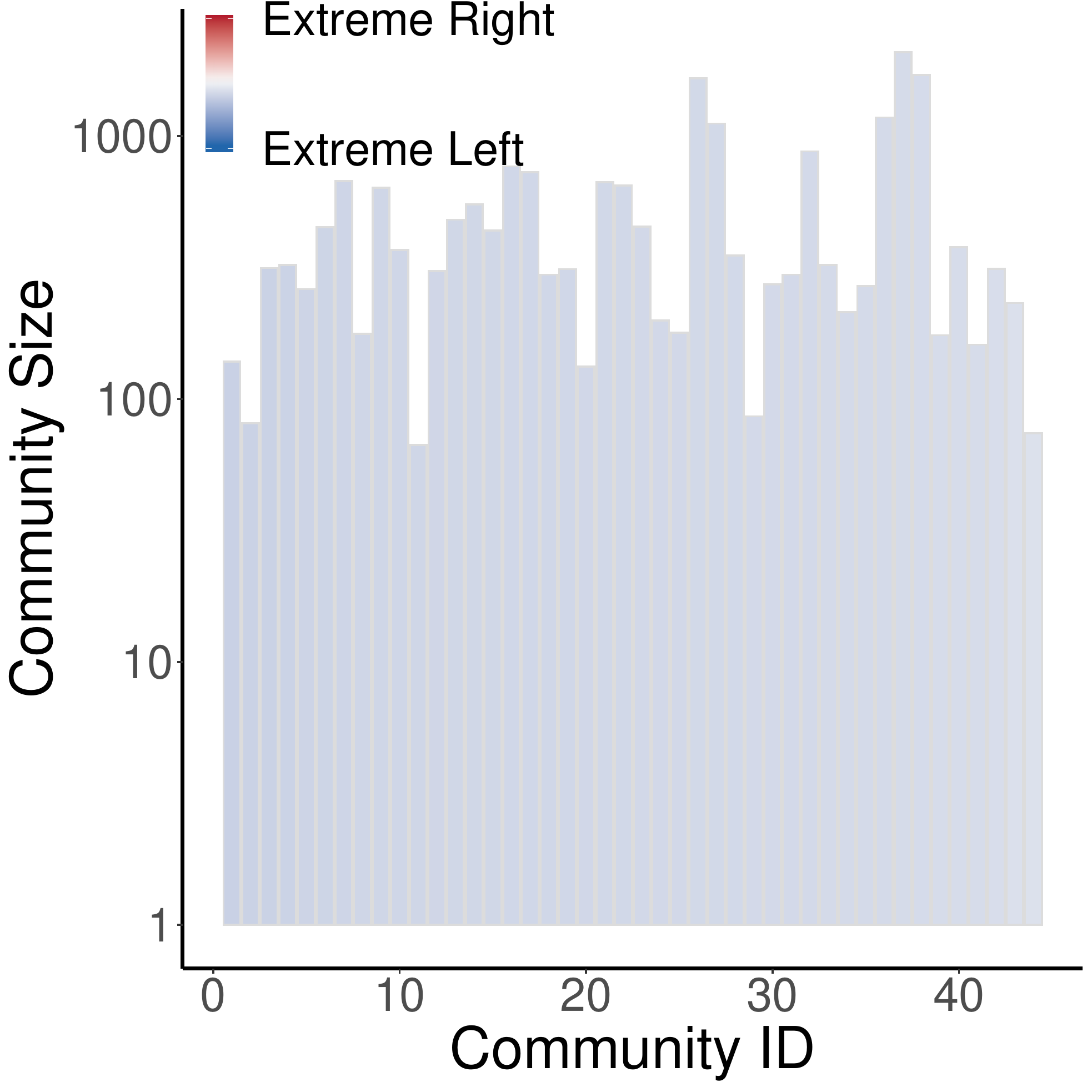}
            \label{fig:reddit_news_SIR}
        \end{subfigure}
                \centering
        \begin{subfigure}[t]{0.23\textwidth}   
            \centering 
            \includegraphics[width=\textwidth]{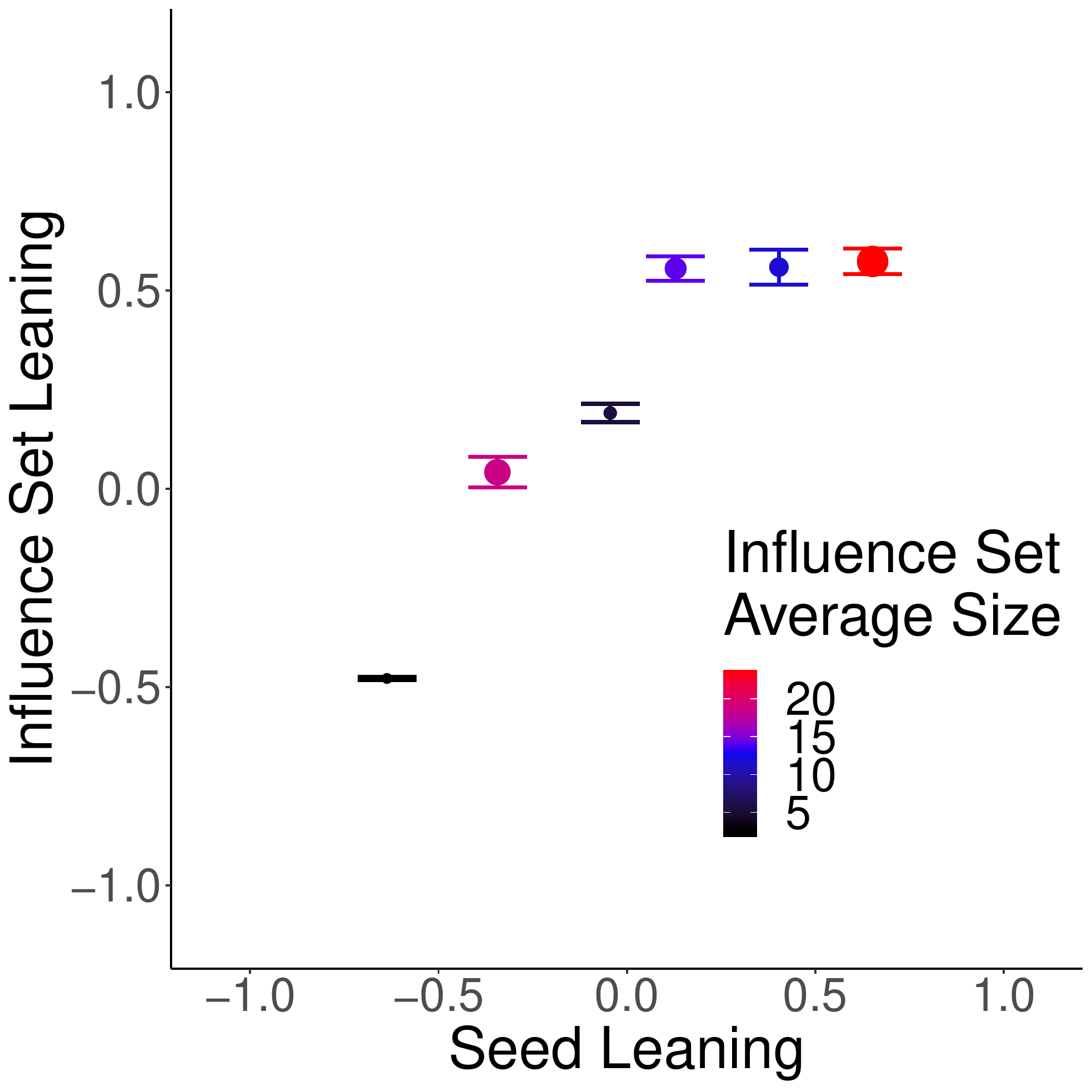}
            \caption{Facebook}
            \label{fig:fb_news_comm}
        \end{subfigure}
        \hfill
        \begin{subfigure}[t]{0.23\textwidth}   
            \centering 
            \includegraphics[width=\textwidth]{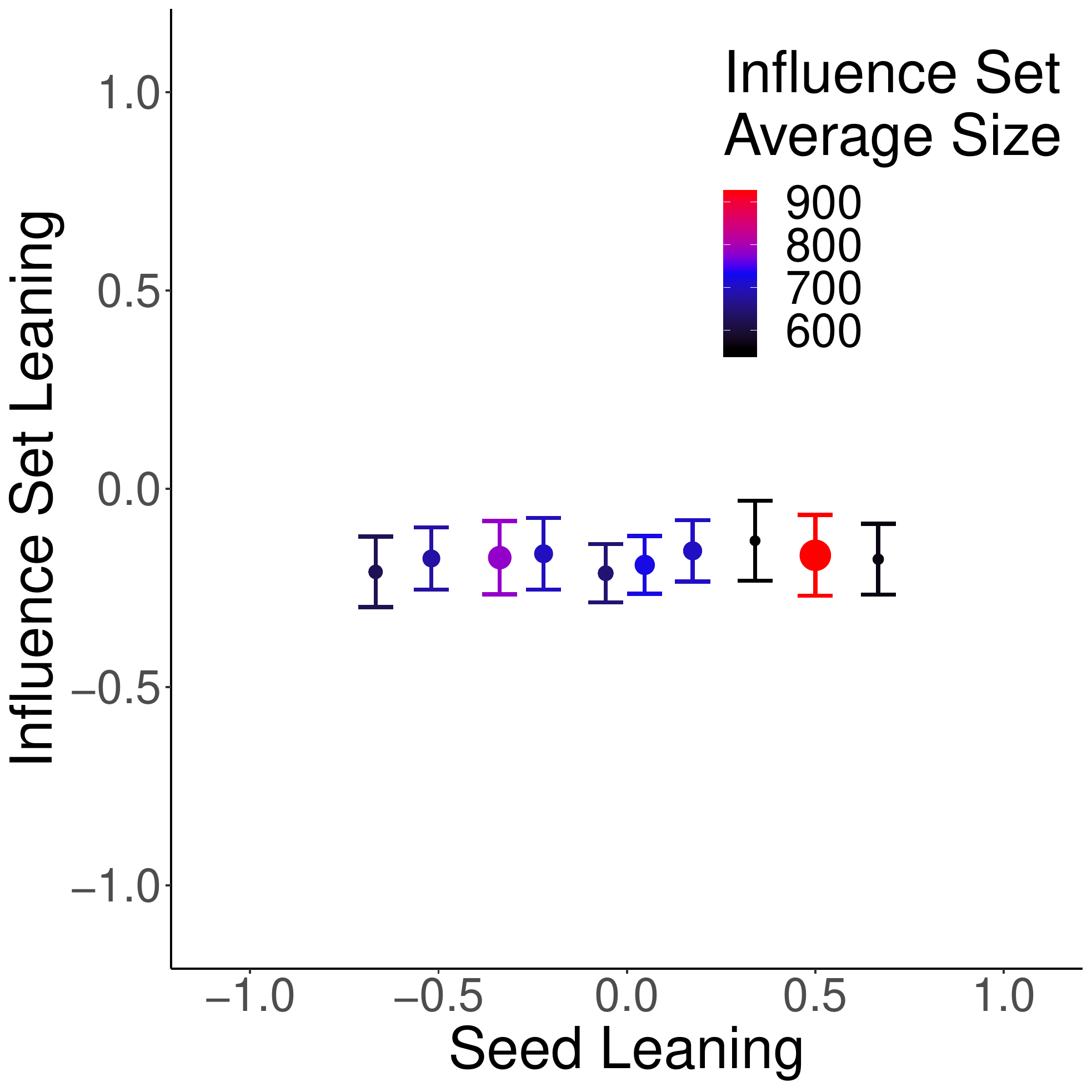}
            \caption{Reddit}
            \label{fig:reddit_news_comm}
        \end{subfigure}%
        \caption{Direct comparison of news consumption on Facebook (left column) and Reddit (right column). 
        Joint distribution of the leaning of users $x$ and the average leaning of their nearest-neighbor $x^{N}$ (top row),  size and average leaning of communities detected in the interaction networks (middle row), and average leaning $\av{\mu(x)}$ of the influence  sets reached by users with leaning $x$, by running SIR dynamics (bottom row) with parameters $\beta=0.05\av{k}$ for panel (a) and $\beta=0.006\av{k}$ for panel (b) and $\nu=0.2$ for both.         \label{fig:news}}
\end{figure}

Fig. \ref{fig:news} shows a direct comparison of news consumption on Facebook and Reddit along the metrics used in the previous Sections to quantify the presence of echo-chambers: i) the correlation between the leaning of a user $x$ and the average leaning of neighbors $x^N$ (top row), ii) the average leaning of communities detected in the networks (middle row), and iii) the average leaning $\av{\mu(x)}$ of the influence  sets reached by users with leaning $x$, by running SIR dynamics (bottom row).  
One can see that all three measures confirm the picture obtained for other data sets: 
On Facebook, we observe a clear separation among users depending on their leaning, while on Reddit users' leanings are more homogeneous and show only one peak. In the latter social media, even users displaying a more extreme leaning (noticeable in the marginal histogram of Figure \ref{fig:news} panel b top row) tend to interact with the majority. Moreover, on Facebook the leaning of the seed user has an effect on who the final recipients of the information are, therefore indicating the presence of echo-chambers. On Reddit this effect is absent.

\section*{Conclusions}

The presence and effects of echo chambers on online social media is a widely debated topic that has profound implications on the way we consume information online and form our opinions.
The wide availability of content combined with confirmation bias and news feed algorithms may foster the emergence of groups of users around a shared narrative. 
Furthermore, the similarity of interests may exacerbate polarization and reinforce existing users tendencies and attitudes. 
To shed light on this issue, in this paper, we introduced an operational definition aimed at identifying echo chambers.
We performed a massive comparative analysis on more than 1B pieces of contents produced by 1M users on four social media platforms: Facebook, Twitter, Reddit, and Gab. 
The proposed method quantifies the presence of echo-chambers along two main dimensions: ($i$) homophily in the interaction networks, and ($ii$) bias in the information diffusion toward likely-minded peers.  
Our results show peculiar differences across social media: while Facebook and Twitter are dominated by echo chambers in all the observed dataset, Reddit and Gab are not.
Furthermore, we perform a direct comparison of news consumption on both Reddit and Facebook.
We found support for the hypothesis that platforms organized around social network and with news feed algorithms which take into account users' preferences foster the emergence of echo-chambers.

\section*{Materials and Methods}
\label{sec:met}

\subsection*{Labelling of media sources}

\begin{figure*}[ht!]
    \centering
    \includegraphics[width=\textwidth]{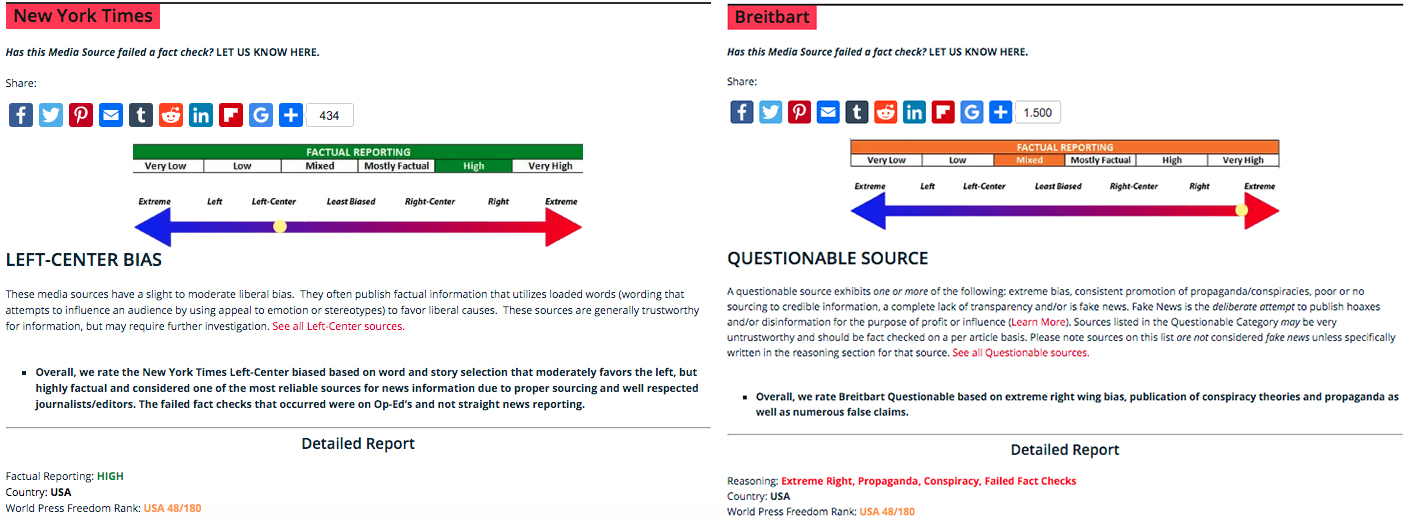}
    \caption{Example of two news sources, namely New York Time and Breitbart, classified on mediabiasfactcheck.org. Notice that, although Breitbart is labeled as "Questionnable", a explicit leaning appears in its description.}
    \label{fig:nyt_vs_breitbart}
\end{figure*}

\begin{figure}[tbp]
\centering
\includegraphics[width=0.48\textwidth]{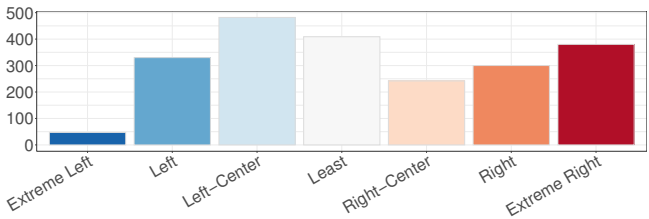}
\caption{Distribution of the leanings assigned to each source, ranging from Extreme Left (numerical value: -1, colored in blue) to Extreme Right (numerical value: +1, colored in red). }
\label{fig:mbfc_leaning}
\end{figure}

The labeling of news outlets is based on the information reported by Media Bias/Fact Check (MBFC~\footnote{\url{https://mediabiasfactcheck.com}}), an independent fact-checking organization that rates news outlets on the base of the reliability and of the political bias of the contents they produce and share.
The website provides the political bias related to a wide range of media outlets.
The labeling provided by MBFC, retrieved in June 2019, ranges from Extreme Left to Extreme Right for what concerns the political bias. 
Certain media outlets are instead classified as `questionable' sources or `conspiracy-pseudoscience' sources if they tend to publish misinformation or false contents.
However, most of the news outlets without an explicit political label reported by MBFC actually have a political bias (e.g., breitbart) that is reported in their description, as shown in Figure~\ref{fig:nyt_vs_breitbart}.
These media outlets often have a political bias that is classified as extreme (either left or right). 
Considering the importance of including such media outlets in our analysis, we manually reported their classification from the description provided by MBFC, thus adding \num{468} outlets to the pool of \num{1722} news outlets that already have a clear political label.
The total number of media outlets for which we have a political label is \num{2190} and the overall leaning is summarized in Figure~\ref{fig:mbfc_leaning}.

\subsection*{Empirical data sets}

Here we report details on data collection for different social media, summarized in Table \ref{tab:data}.

\begin{table*}[tb]
\caption{For each data set, we report: the starting date of collection $T_0$, time span $T$ expressed in days (d) or years (y), number of unique contents $C$, number of users $N$, coverage  $n_c$ (fraction of users with classified leaning), size of the giant component $G$ and average node degree $\av{k}$.  }
\label{tab:data}
\begin{ruledtabular}
    \begin{tabular}{l l c r r r r r r}
  Media  &  Data set &  $T_0$&  $T$ &  $C$ &    $N$ &  $n_c$ & $G$ & $\av{k}$ \\ \hline 
 \multirow{3}{*}{Twitter}  & Gun control & 06/2016 &  14 d & $19$ M & \num{7506}& 0.93 & 3964 & 798 \\
                                       & Obamacare & 06/2016 & 7 d & $34$ M	& \num{8773} & 0.90 & 8703 & 1405\\
                                        & Abortion & 06/2016	& 7 d & $34$ M	& \num{3995} & 0.95 & 798 & 478 \\ \hline 
 \multirow{3}{*}{Facebook} & Sci/Cons & 01/2010	& 5 y & \num{75172} & \num{183378}	& 1.00 & 181960 & 228	\\
                                        & Vaccines & 01/2010 & 7 y & \num{94776}	 & \num{221758}	&  1.00 & 220275 & 419	\\
                                       & News 	& 01/2010	& 6 y & \num{15540}	& \num{38663}	 & 1.00 & 38594 & 700	\\   \hline
 \multirow{3}{*}{Reddit}   & Politics & 01/2017	& 1 y & \num{353864} & \num{240455}	&  0.15 & 240455 & 9	 \\
                                           & The Donald 	& 01/2017 & 1 y &  \num{1.234} M	 & \num{138617}	& 0.16 & 138617 & 31	\\
                                       & News & 01/2017	 & 1 y & \num{723235}	 & \num{179549}	& 0.20 & 179549 &	3 \\   \hline 
  \multirow{1}{*}{Gab} & Gab & 11/2017	& 1 y & $13$ M	& \num{165162}	& 0.13 & 20701 & 328 \\ 
    \end{tabular}
\end{ruledtabular}
  \label{tab:crit_exp}
\end{table*}


\subsubsection*{Twitter}

\spara{Gun control.} 
We consider $C=19$M tweets spanning 14 days in June 2016, produced by $N=7506$ users. 
We reconstruct a directed follow network formed by $E=\num{1053275}$ directed edges. 
The largest weakly connected component includes more than $99\%$ of nodes. 
We identify the individual leaning of $N_c=\num{6994}$ users. 

\spara{Obamacare.}
We consider $C=34$M tweets spanning 7 days in June 2016, produced by $N=8773$ users. 
We reconstruct a directed follow network formed by $E=\num{3797871}$ directed edges. 
The largest weakly connected component includes more than $99\%$ of nodes. 
We identify the individual leaning of $N_c=\num{7899}$ users. 

\spara{Abortion.}
We consider $C=34$M tweets spanning 7 days in June 2016, produced by $N=3995$ users. 
We reconstruct a directed follow network formed by $E=\num{2330276}$ directed edges. 
The largest weakly connected component includes more than $99\%$ of nodes. 
We identify the individual leaning of $N_c=\num{3809}$ users.

\subsubsection*{Facebook}

\spara{Science and Conspiracy.}
The dataset was built by downloading posts of selected Facebook pages divided into two groups, namely conspiracy news and science news.
Conspiracy pages were selected based on their name, their self description and with the aid of debunking pages.
The selection process was iterated until convergence among annotators.
The dataset, that includes post from pages and comments to such posts, was created by using Facebook Graph API and has previously been explored~\cite{bessi2016users}.
We consider \num{75172} posts by 73 pages categorized in Science (34) and Conspiracy (39) that involve $N=\num{183378}$ active users (at least 1 like and 1 comments) that co-commented \num{20807976} times.
The largest connected component of the co-commenting network has $G = \num{181960}$ nodes and $E = \num{20807491}$ links.

\spara{Vaccines.}
The dataset was generated in three steps: first a search for pages containing the keywords vaccine, vaccines, or vaccination was made.
Then the raw outcome was cleaned from spurious pages.
Finally all the posts and comments of selected pages were downloaded and pages were manually classified in Pro-Vax and Anti-Vax groups.
The dataset was created by using Facebook Graph API and has previously been explored~\cite{schmidt2018polarization}.
Thus, we consider \num{94776} posts by 243 pages categorized in Pro-Vax (145) and Anti-Vax (98) that involve \num{221758} active users (at least 1 like and 1 comment) that co-commented \num{46198446} times.  The largest connected components of the co-commenting network has $N = \num{220275}$ nodes and $E = \num{46193632}$ links.

\spara{News.}
To build this dataset, a set of Facebook pages of news outlets listed by the Europe Media Monitor was identified as first step.
By using the Facebook Graph API, all the posts and comments related to these pages in the periods between 2010-2015 were downloaded.
Facebook pages are labelled according to the annotation provided by mediabiasfactcheck.org.
The dataset without annotations and has previously been explored~\cite{schmidt2017anatomy}.
We consider \num{15540} posts by 180 pages categorized from Left to Right (Left (12), Left-Center (80), Least-Biased (42), Right-Center (33), Right (13)),
38663 active users ($\geq$ 3 likes and 3 comments) that co-commented \num{13525230} times.
The largest connected component of the co-interaction network has $G = \num{38594}$ nodes and $E = \num{13525119}$ links.

\subsubsection*{Reddit}

\spara{Politics.} 
We consider \num{353864} comments and submissions posted on the subreddit \emph{politics} in the year 2017.
From comments a submissions we reconstructed a directed network formed by $N=\num{240455}$ users and $E=\num{5030565}$ directed edges. 
The largest weakly connected component includes more than $0.99\%$ of nodes. 
We identify the individual leaning of $N_c=\num{37148}$ users. 

\spara{The Donald.}
We consider \num{1.234}M comments and submissions posted on the subreddit \emph{The\_Donald} in the year 2017. 
From comments a submissions we reconstructed a directed network formed by $N=\num{138617}$ users and $E=\num{5025290}$ directed edges. 
The largest weakly connected component includes more than $0.99\%$ of nodes. 
We identify the individual leaning of $N_c=\num{21905}$ users. 

\spara{News.}
We consider \num{723235} comments and submissions posted on the subreddit \emph{news} in the year 2017.
From comments a submissions we reconstructed a directed network formed by $N=\num{179549}$ users and $E=\num{1070589}$ directed edges. 
The largest weakly connected component includes more than $0.99\%$ of nodes. 
We identify the individual leaning of $N_c=\num{36875}$ users. 

\subsubsection*{Gab}

The dataset, downloaded from \url{https://files.pushshift.io/gab}, spans from the first Gab post (occurred in 2016) to the late 2018 and it includes data regarding post-reply relationships, number of upvotes of posts, repost or replies and their timestamps. 
We selected all the contents (post, reply, quote) from 11/2017 to 10/2018, that is $C=\num{13580937}$ unique pieces of content created by $N=\num{165162}$ unique users.
We consider all the post that have a link to an external source, for an amount of \num{3302621} posts (excluding youtube links).
By extracting the domain from each link we obtain \num{75436} unique domains.
In this set, \num{1650} unique domains for a total of \num{1454502} URLs ($44\%$) were labelled in the MBFC database.
We were able to compute the political leaning of $N_c = \num{31286}$ users.
We also reconstructed the interaction network using co-commenting as a proxy.
The largest connected component includes $G=\num{20701}$ nodes, about the 66\% of the users with leaning, $E=\num{8273412}$ edges.

\bibliographystyle{unsrt}
\bibliography{references}
\clearpage

\input{SI.tex}
\end{document}

%% file: SI.tex
\onecolumngrid
\begin{center}

    \textsc{\Large{Supplementary Information \\Echo Chambers on Social Media:  A comparative analysis}}
\end{center}
Here we show additional results not shown in the main paper: additional data sets in Section \ref{sec:emp} and additional results for the SIR dynamics run with different parameters in Section \ref{sec:sir}

\section{Additional data sets}
\label{sec:emp}

In this section we report the results obtained for other four data sets not shown in the main paper, namely ``Science and Conspiracy" (Facebook), ``Gun control" (Twitter), ``Obamacare" (Twitter) and `The Donald" (Reddit). 
The techniques and the pipeline is the same used for the datasets analyzed in the main paper.

\subsection{Science and Conspiracy}

\begin{figure*}[ht]
\centering
        \begin{subfigure}[t]{0.33\textwidth}   
            \centering 
            \includegraphics[width=\textwidth]{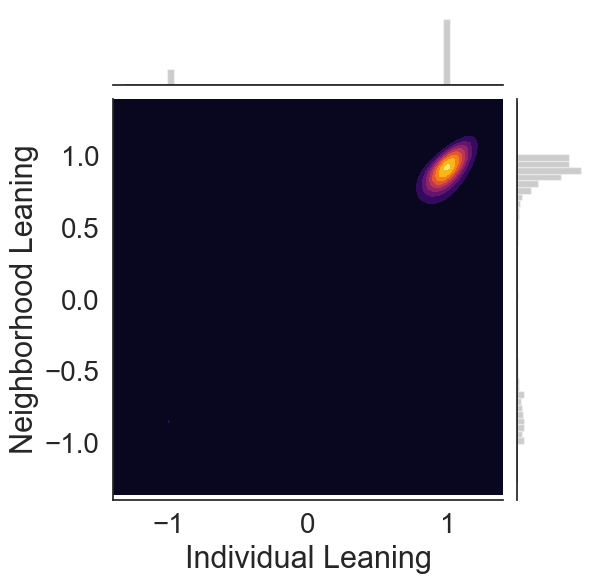}
            \caption{}
        \end{subfigure}
        \begin{subfigure}[t]{0.33\textwidth}   
            \centering 
            \includegraphics[width=\textwidth]{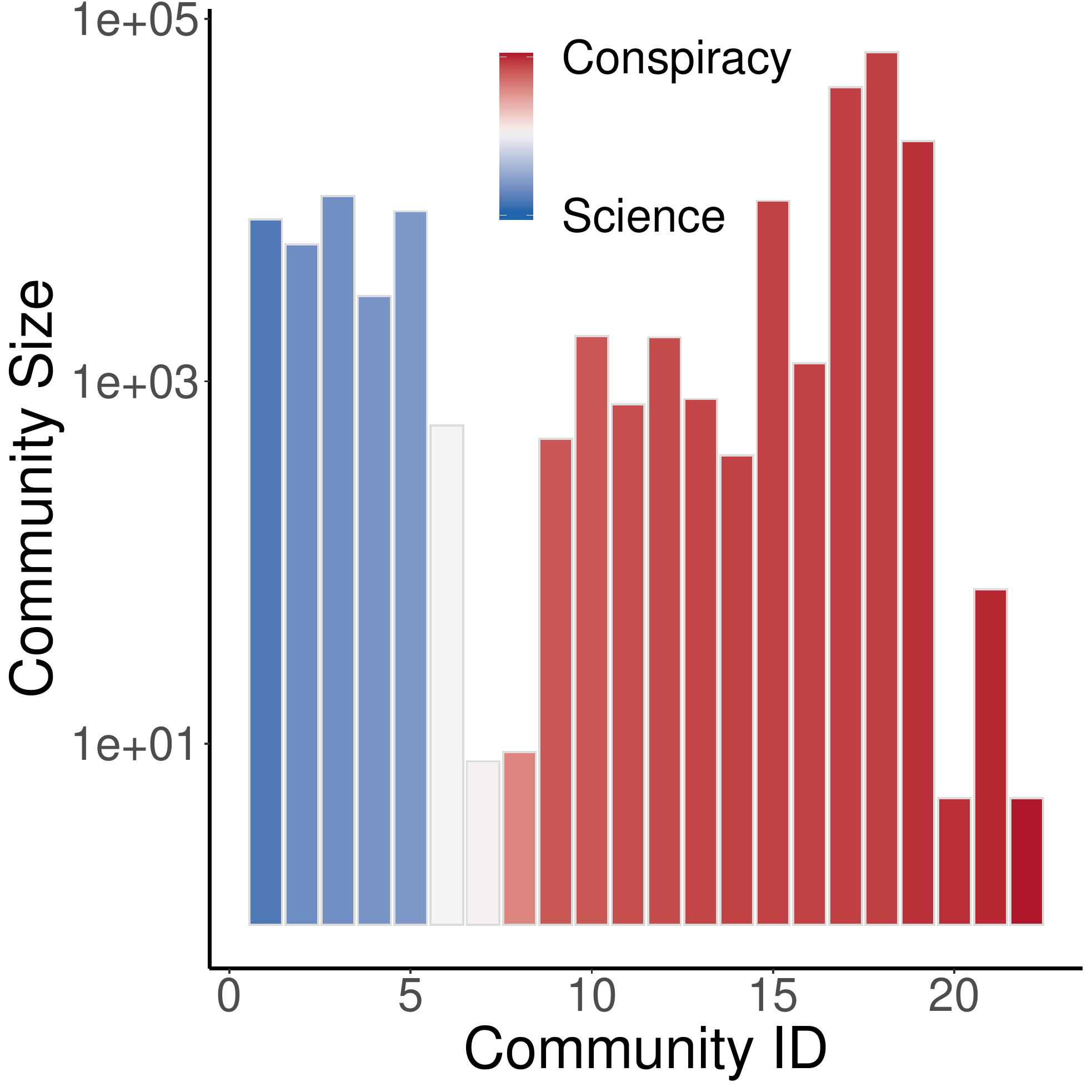}
            \caption{}
        \end{subfigure}
        \begin{subfigure}[t]{0.33\textwidth}   
            \centering 
            \includegraphics[width=\textwidth]{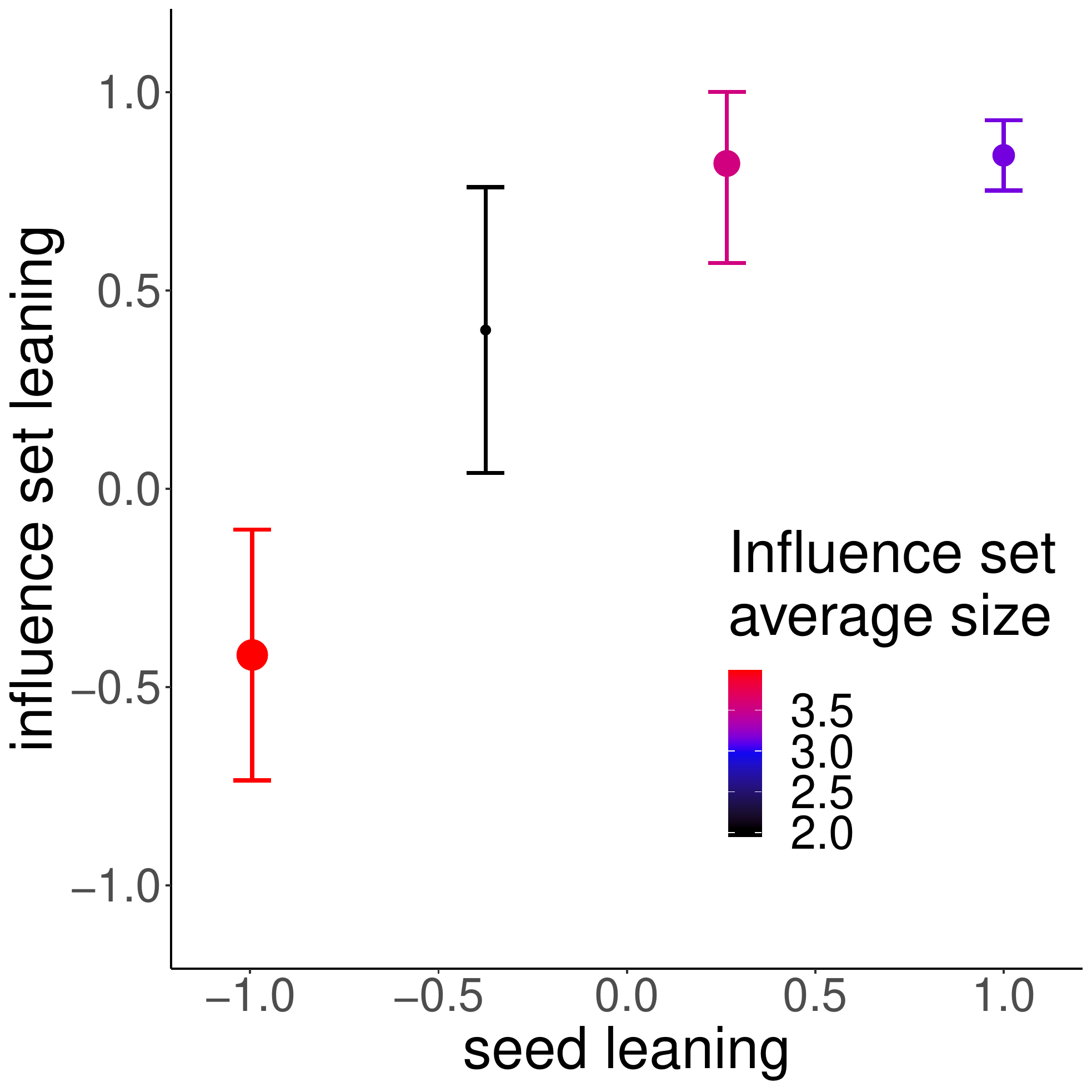}
            \caption{}
        \end{subfigure}
        \begin{subfigure}[t]{0.33\textwidth}   
            \centering 
            \includegraphics[width=\textwidth]{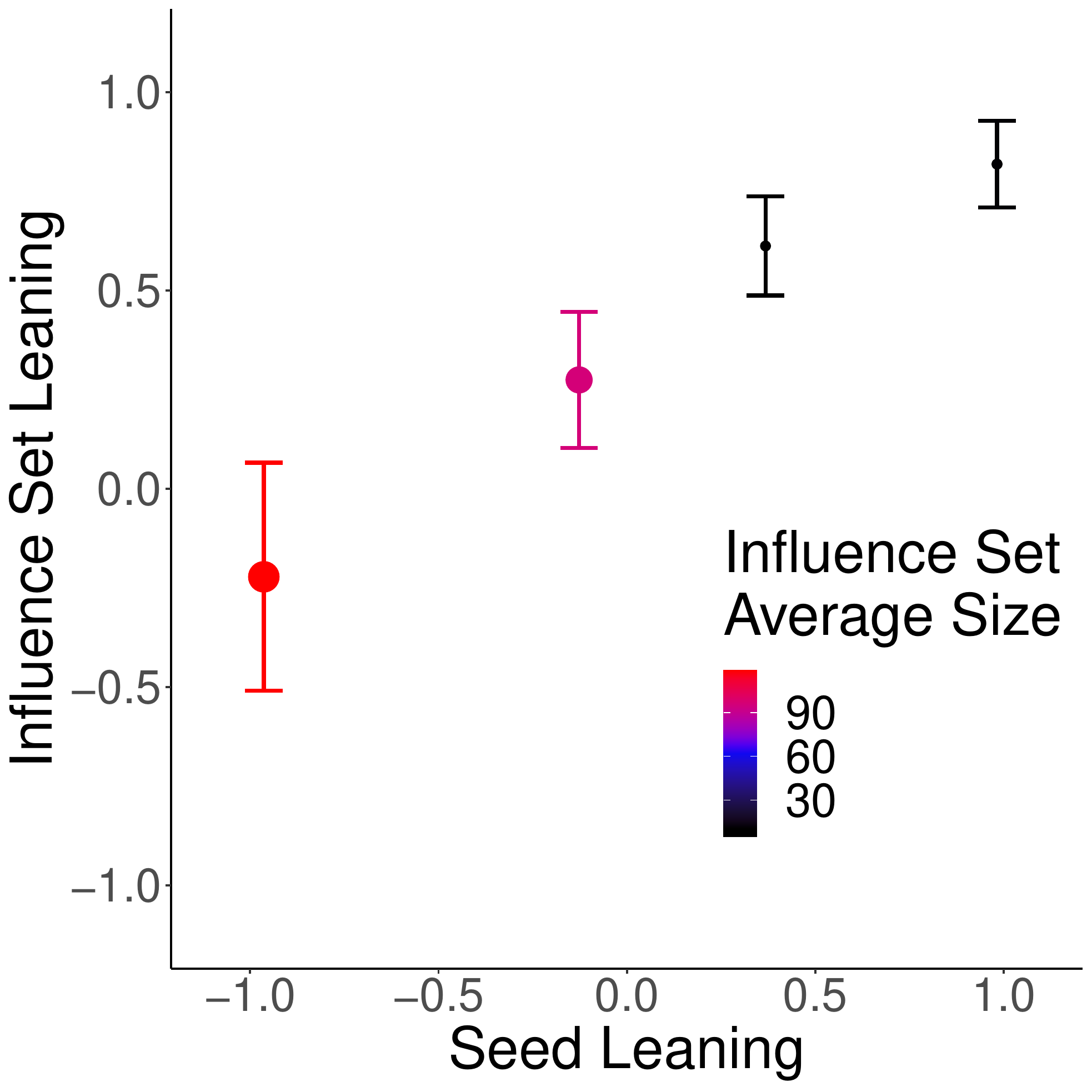}
            \caption{}
        \end{subfigure}
        \caption{Science vs Conspiracy.  Panel (a): Individual leaning versus neighborhood leaning. Panel (b): Community detection. Panel (c) and (d): average leaning $\langle \mu(x) \rangle$ of the influence sets reached by users with leaning $x$, for infection probability $\beta=0.01 \langle k \rangle^{-1}$ and $\beta=0.02 \langle k \rangle^{-1}$, respectively, where $ \langle k \rangle$ is the average degree of the network.}
        \label{scicos}
\end{figure*}

Figure~\ref{scicos} displays the results obtained for the Facebook dataset called ``Science and Conspiracy", described in Materials and Methods of the main paper. 
Panel (a) shows the joint distribution of the leaning of users, $x$, against the average leaning of their neighborhood $X^N$. 
We note that the community referred to as ``Science", to which is associated a leaning of -1, is much smaller than the community called "Conspiracy" and for this reason it is not clearly visible in the density plot but only in the histograms at its margins. Panel (b) shows the size and average leaning of communities detected by the Louvain algorithm. 
 
Panels (c) and (d) show the results of the SIR dynamics: the average leaning $\langle \mu(x) \rangle$ of the influence sets reached by users with leaning $x$, for two different values of the infection probability, while the recovery rate is fixed $\nu = 0.2$.
Size and color of each point is related to the average size of the influence sets.

\subsection{Guncontrol}
\begin{figure*}[ht]
\centering
        \begin{subfigure}[t]{0.33\textwidth}   
            \centering 
            \includegraphics[width=\textwidth]{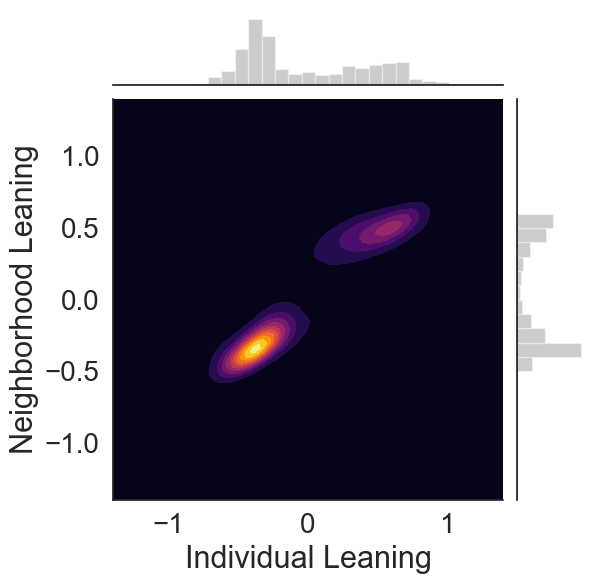}
            \caption{}
        \end{subfigure}
        \begin{subfigure}[t]{0.33\textwidth}   
            \centering 
            \includegraphics[width=\textwidth]{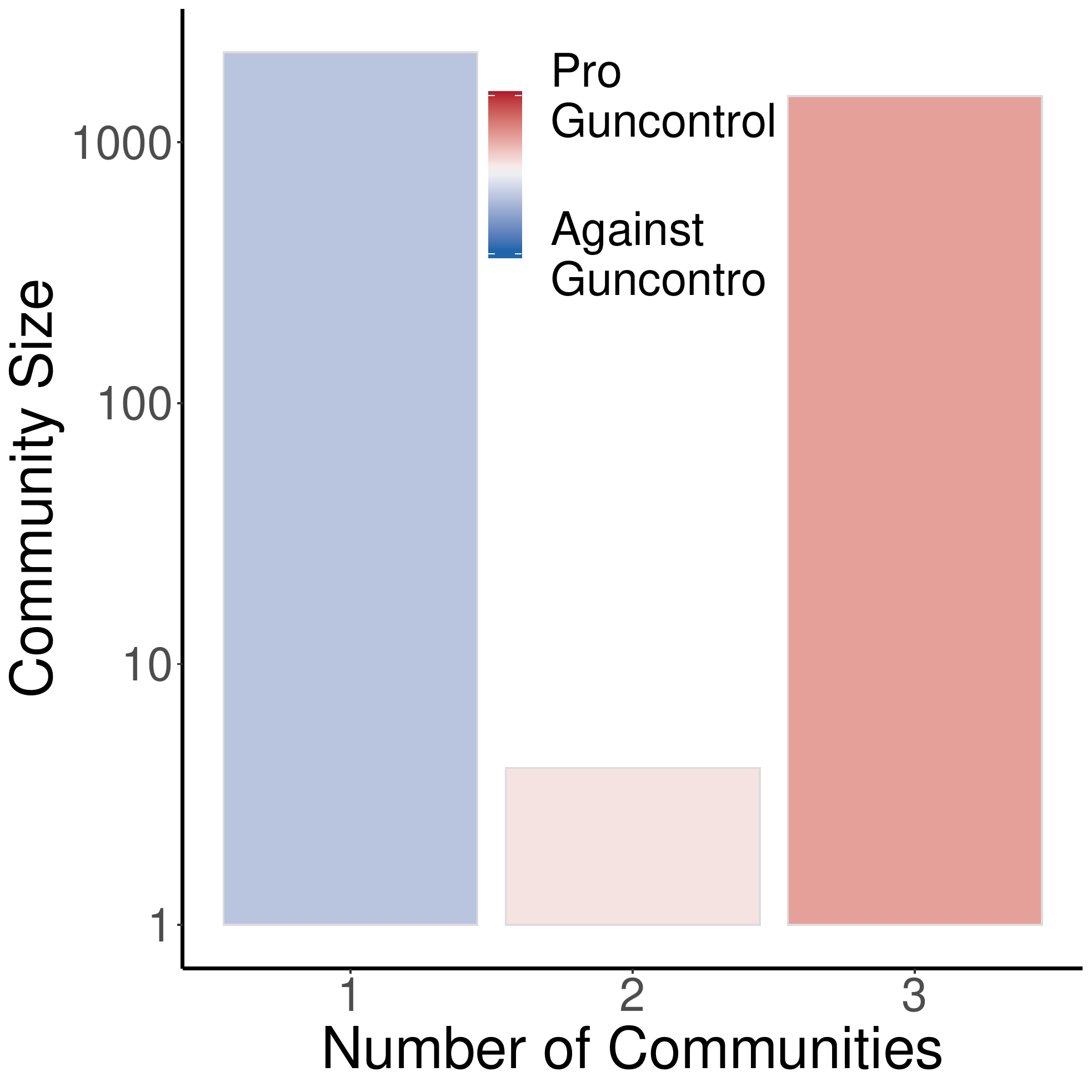}
            \caption{}
        \end{subfigure}
        \begin{subfigure}[t]{0.33\textwidth}   
            \centering 
            \includegraphics[width=\textwidth]{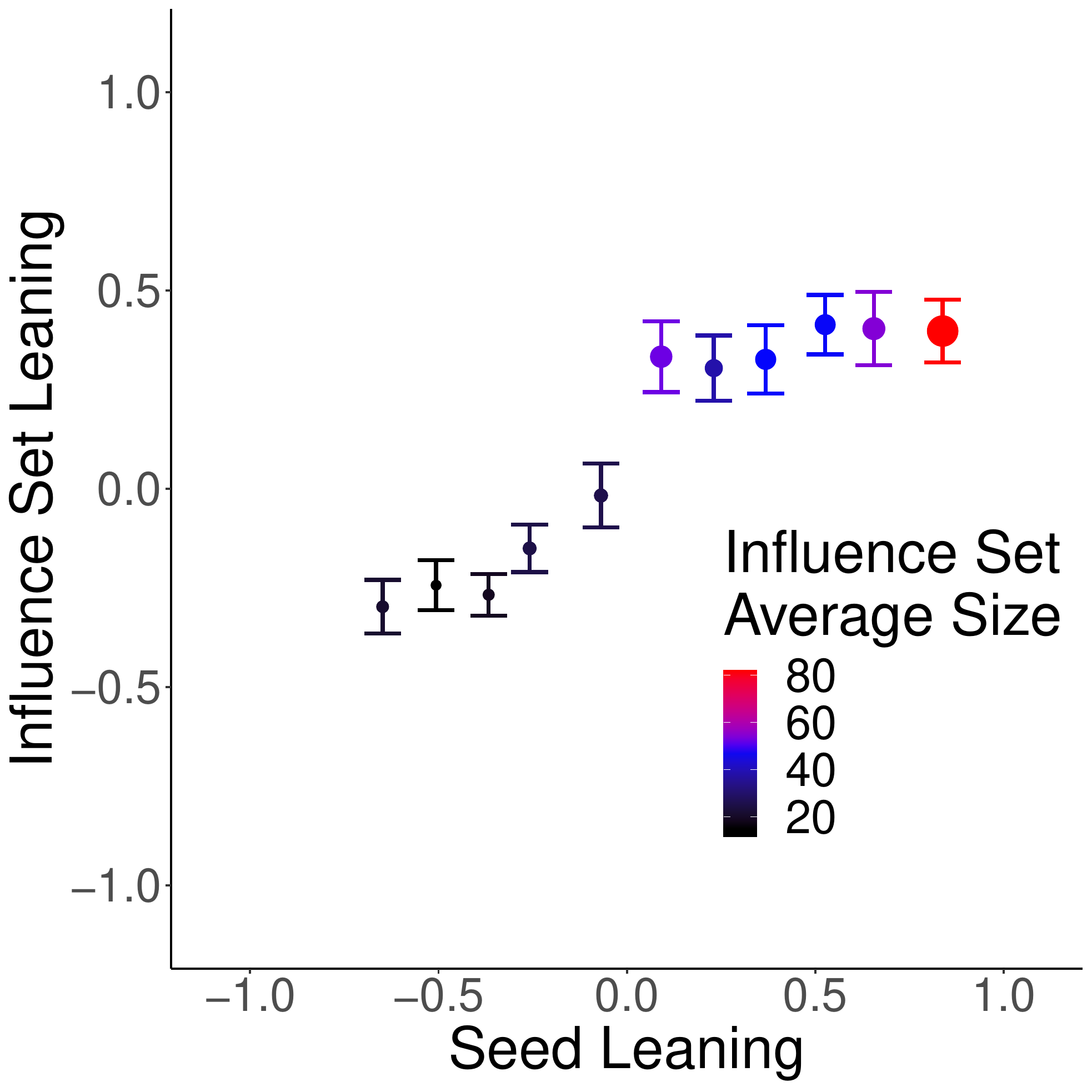}
            \caption{}
        \end{subfigure}
        \begin{subfigure}[t]{0.33\textwidth}   
            \centering 
            \includegraphics[width=\textwidth]{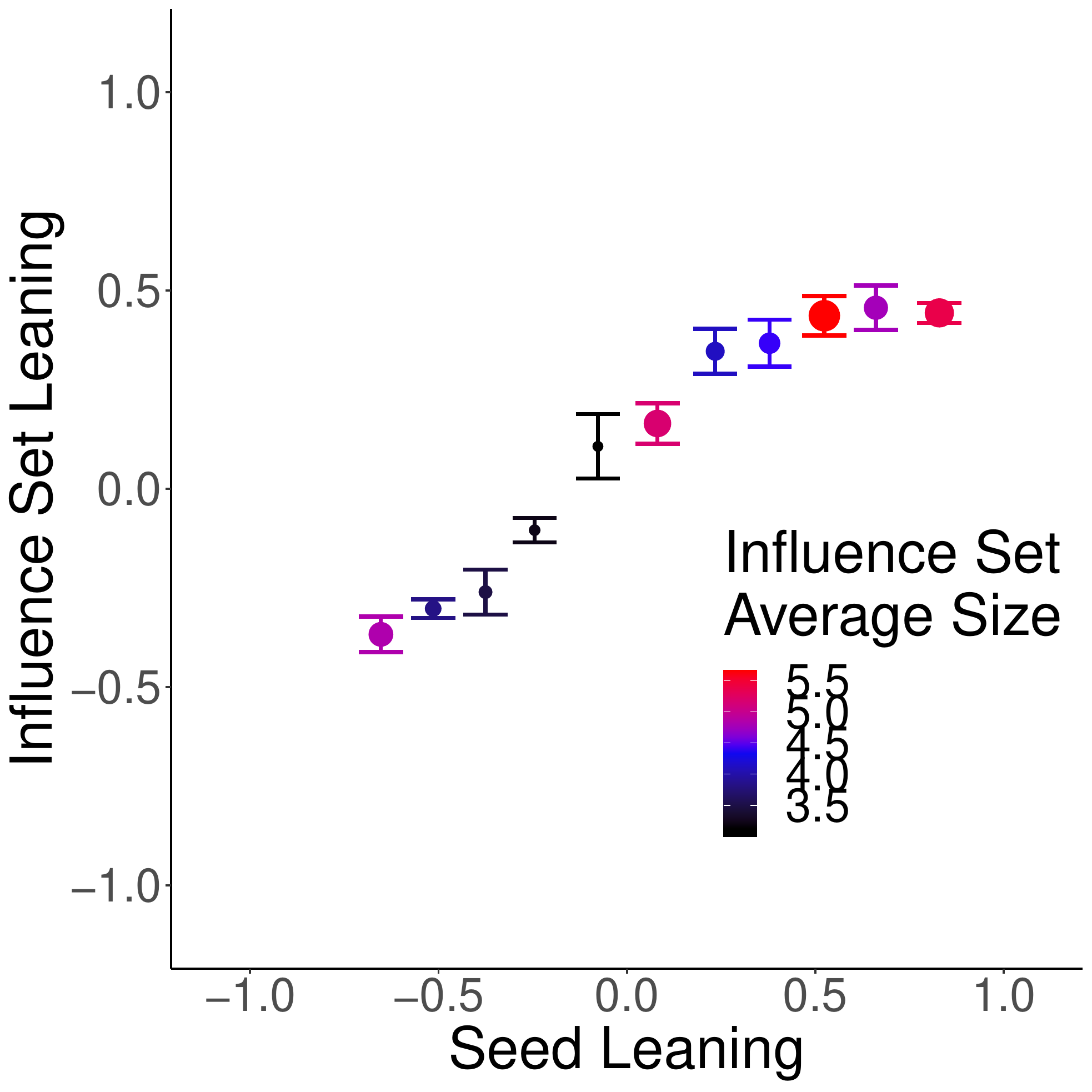}
            \caption{}
        \end{subfigure}
        \caption{Gun control. 
        Panel (a): Individual leaning versus neighborhood leaning. Panel (b): Community detection. Panel (c) and (d): average leaning $\langle \mu(x) \rangle$ of the influence sets reached by users with leaning $x$, for infection probability $\beta=0.1 \langle k \rangle^{-1}$ and $\beta=0.2 \langle k \rangle^{-1}$, respectively, where $ \langle k \rangle$ is the average degree of the network.}
        \label{guncontrol}
\end{figure*}

Figure \ref{guncontrol} shows the results obtained for the Twitter dataset ``Gun control", described in Materials and Methods of the main paper. 
Panel (a) shows the joint distribution of the leaning of users, $x$, against the average leaning of their neighborhood $X^N$, in which two different regions are clearly visible. 
Panel (b) shows the size and average leaning of communities detected by the Louvain algorithm. 

Panels (c) and (d) show the results of the SIR dynamics: the average leaning $\langle \mu(x) \rangle$ of the influence sets reached
by users with leaning $x$, for two different values of the infection probability, while the recovery rate is fixed $\nu = 0.2$.
Size and color of each point is related to the average size of the influence sets.

\subsection{Obamacare}

\begin{figure*}[ht]
\centering
        \begin{subfigure}[t]{0.33\textwidth}   
            \centering 
            \includegraphics[width=\textwidth]{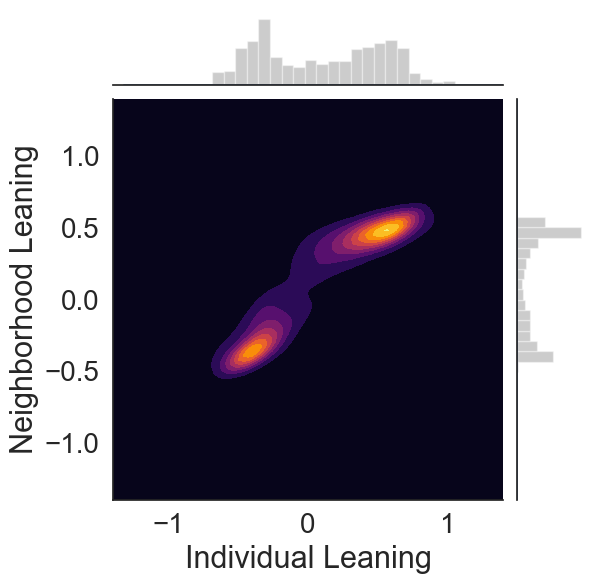}
            \caption{}
        \end{subfigure}        
        0.33
        \begin{subfigure}[t]{0.33\textwidth}   
            \centering 
            \includegraphics[width=\textwidth]{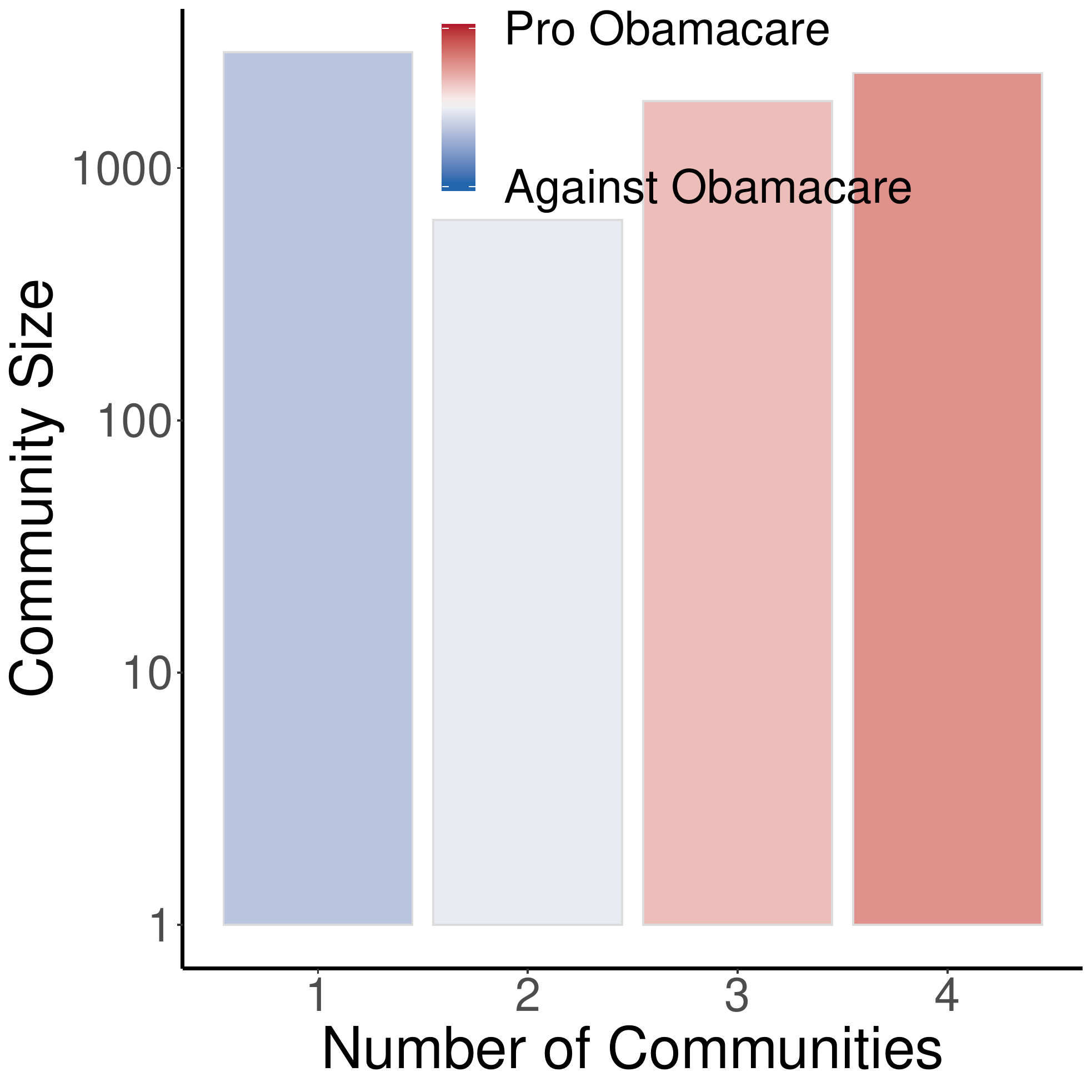}
            \caption{}
        \end{subfigure}
        \begin{subfigure}[t]{0.33\textwidth}   
            \centering 
            \includegraphics[width=\textwidth]{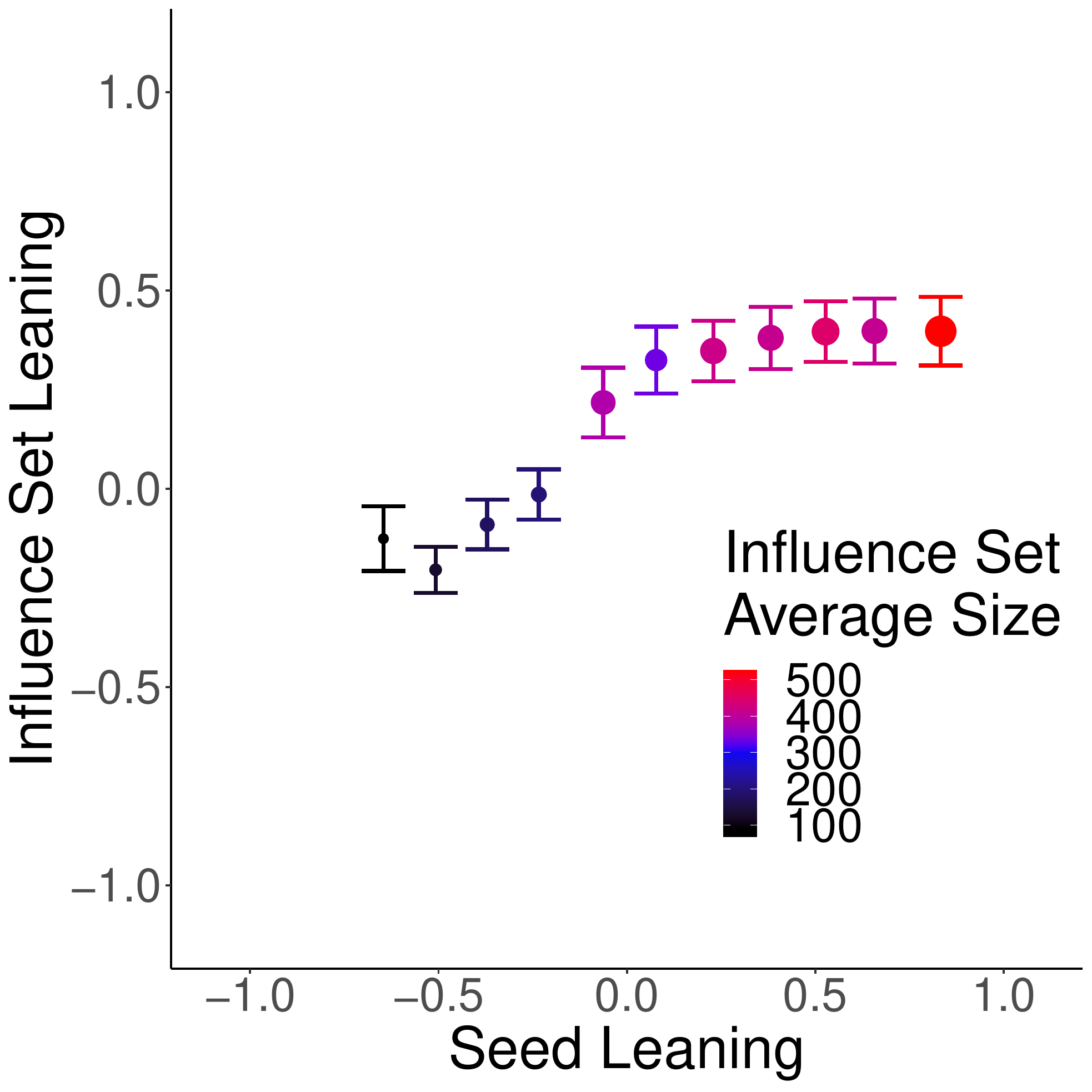}
            \caption{}
        \end{subfigure}
        \begin{subfigure}[t]{0.33\textwidth}   
            \centering 
            \includegraphics[width=\textwidth]{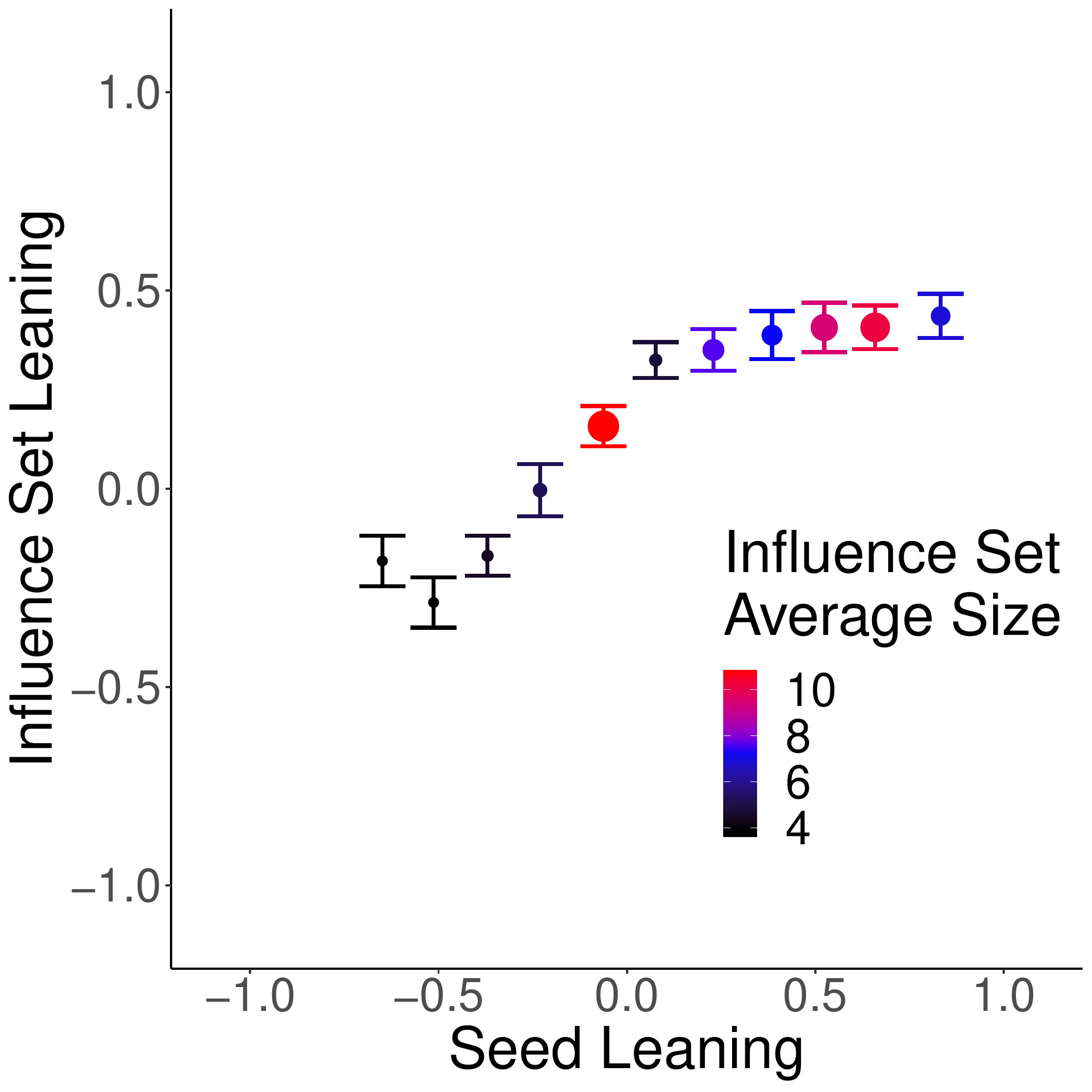}
            \caption{}
        \end{subfigure}
        \caption{Obamacare. Panel (a): Individual leaning versus neighborhood leaning. Panel (b): Community detection. Panel (c) and (d): average leaning $\langle \mu(x) \rangle$ of the influence sets reached by users with leaning $x$, for infection probability $\beta=0.1 \langle k \rangle^{-1}$ and $\beta=0.2 \langle k \rangle^{-1}$, respectively, where $ \langle k \rangle$ is the average degree of the network.}
        \label{obamacare}
\end{figure*}

Figure \ref{obamacare} shows the results obtained for the Twitter dataset referred to as ``Obamacare", described in Materials and Methods of the main paper. 
Panel (a) shows the joint distribution of the leaning of users, $x$, against the average leaning of their neighborhood $X^N$, in which two interconnected regions are clearly visible. 
Panel (b) shows the size and average leaning of communities detected by the Louvain algorithm. 

 Panels (c) and (d) show the results of the SIR dynamics: the average leaning $\langle \mu(x) \rangle$ of the influence sets reached
by users with leaning $x$, for two different values of the infection probability,  while the recovery rate is fixed $\nu = 0.2$.
Size and color of each point is related to the average size of the influence sets.  

\subsection{TheDonald}
\begin{figure*}[ht]
\centering
        \begin{subfigure}[t]{0.33\textwidth}   
            \centering 
            \includegraphics[width=\textwidth]{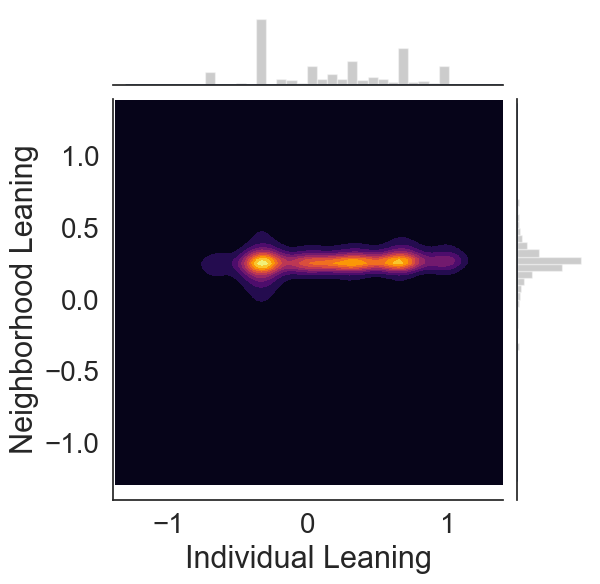}
            \caption{}
        \end{subfigure}
        \begin{subfigure}[t]{0.33\textwidth}   
            \centering 
            \includegraphics[width=\textwidth]{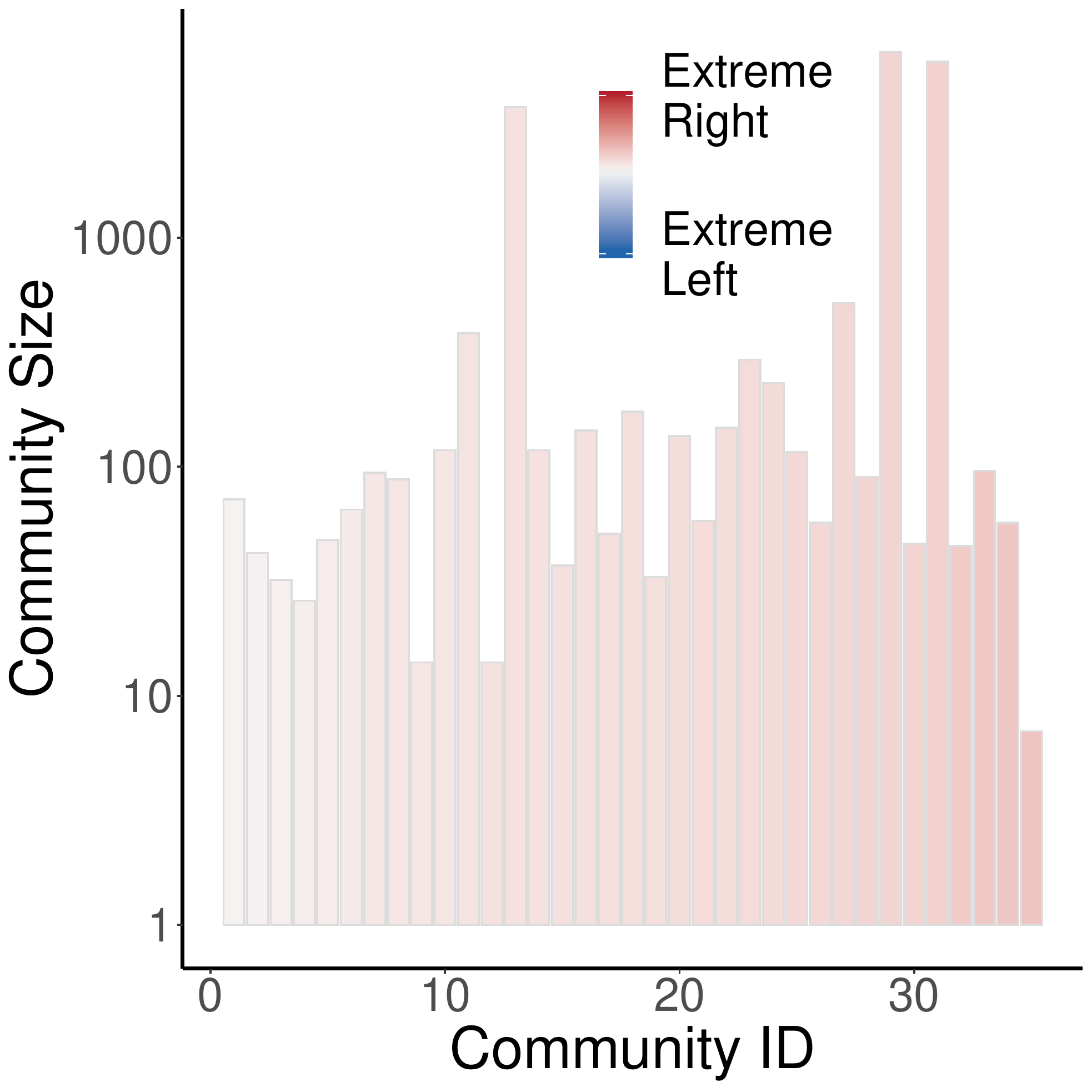}
            \caption{}
        \end{subfigure}
        \begin{subfigure}[t]{0.33\textwidth}   
            \centering 
            \includegraphics[width=\textwidth]{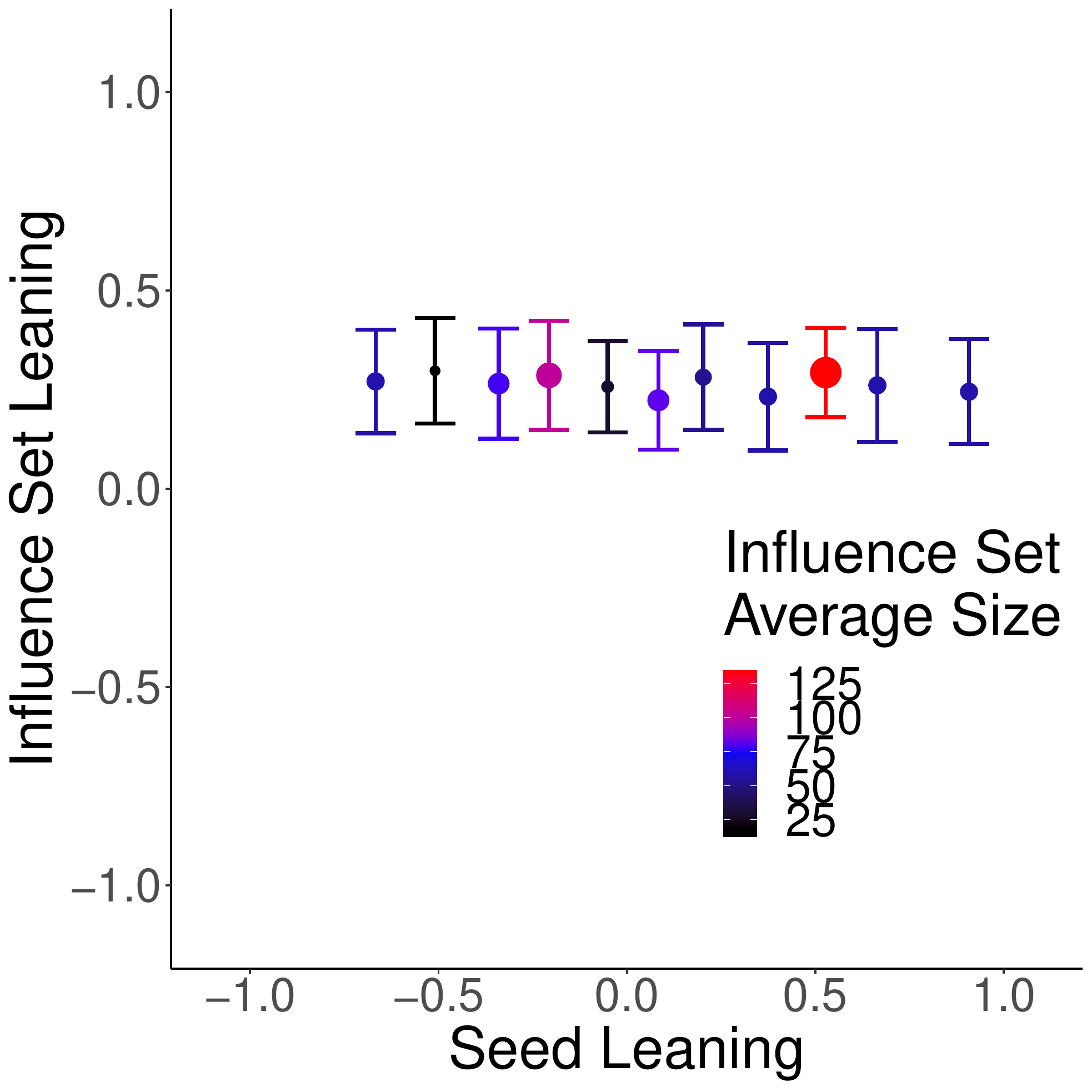}
            \caption{}
        \end{subfigure}
        \begin{subfigure}[t]{0.33\textwidth}   
            \centering 
            \includegraphics[width=\textwidth]{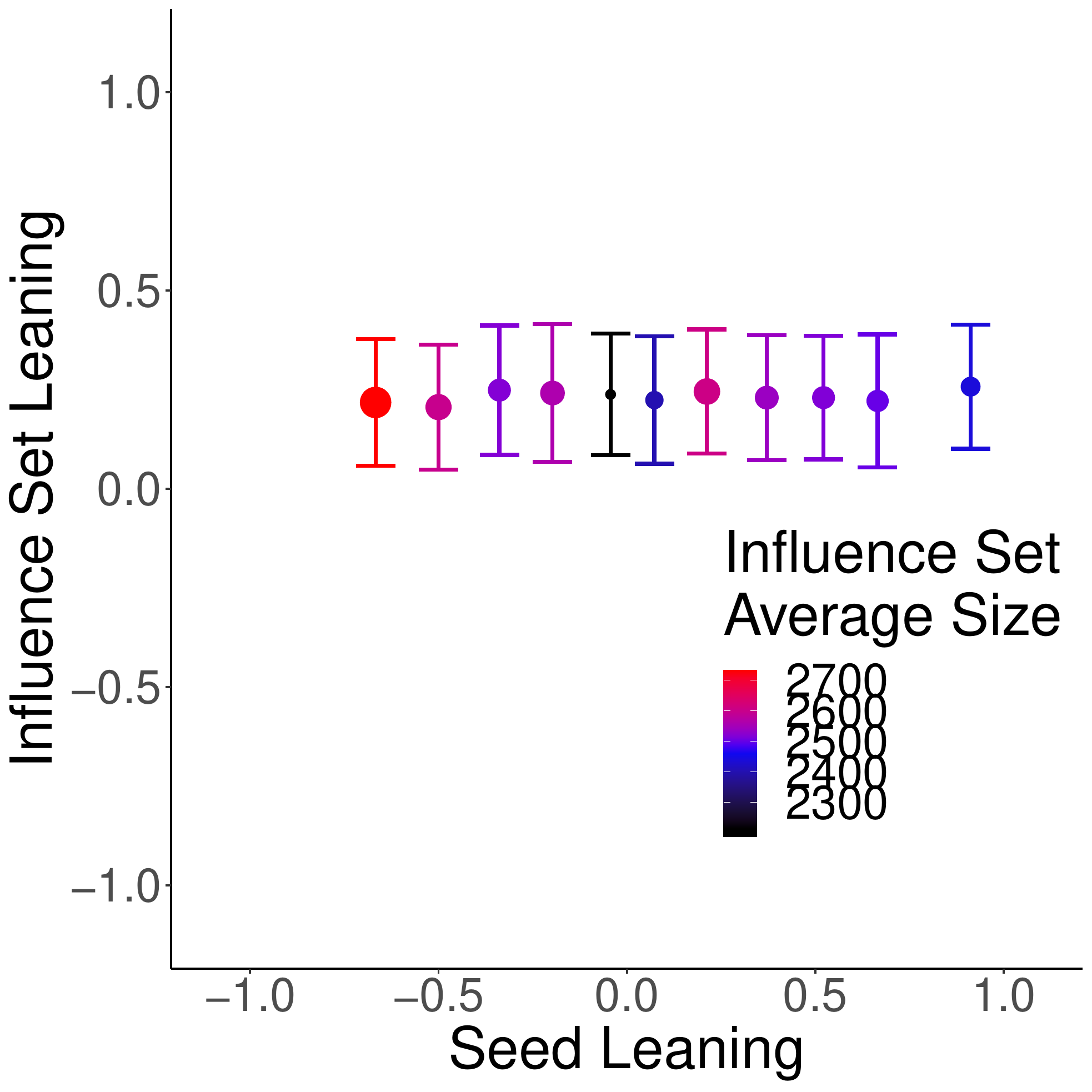}
            \caption{}
        \end{subfigure}
        \caption{The Donald. Panel (a): Individual leaning versus neighborhood leaning. Panel (b): Community detection. Panel (c) and (d): average leaning $\langle \mu(x) \rangle$ of the influence sets reached by users with leaning $x$, for infection probability $\beta=0.0067 \langle k \rangle^{-1}$ and $\beta=0.013 \langle k \rangle^{-1}$, respectively, where $ \langle k \rangle$ is the average degree of the network.}
        \label{thedonald}
\end{figure*}
Figure \ref{thedonald} shows the results obtained for the Reddit dataset ``The Donald", described in Materials and Methods of the main paper. 
Panel (a) displays the joint distribution of the leaning of users, $x$, against the average leaning of their neighborhood $X^N$, showing a unique region spanning most of the x-axis and concentrated on the values around 0.25 on the y-axis. Such a region is also characterized by few  peaks of leaning (spanning mainly from Center to Extreme Right) that are displayed in the histogram on the top margin. 
Panel (b) shows the size and average leaning of communities detected by the Louvain algorithm. 

 Panels (c) and (d) show the results of the SIR dynamics: the average leaning $\langle \mu(x) \rangle$ of the influence sets reached
by users with leaning $x$, for two different values of the infection probability, while the recovery rate is fixed $\nu = 0.2$.
Size and color of each point is related to the average size of the influence sets.

\newpage
\clearpage

\section{Robustness of the SIR dynamics}
\label{sec:sir}

In this section, we provide additional results for the SIR dynamics run with different parameters on the 6 data sets considered in the main paper, namely ``Abortion" on Twitter, ``Politics" and ``News" on Reddit, ``Vaccines" and ``News" on Facebook, and Gab.

\begin{figure*}[ht]
\centering
        \begin{subfigure}[t]{0.3\textwidth}
            \centering 
            \includegraphics[width=\textwidth]{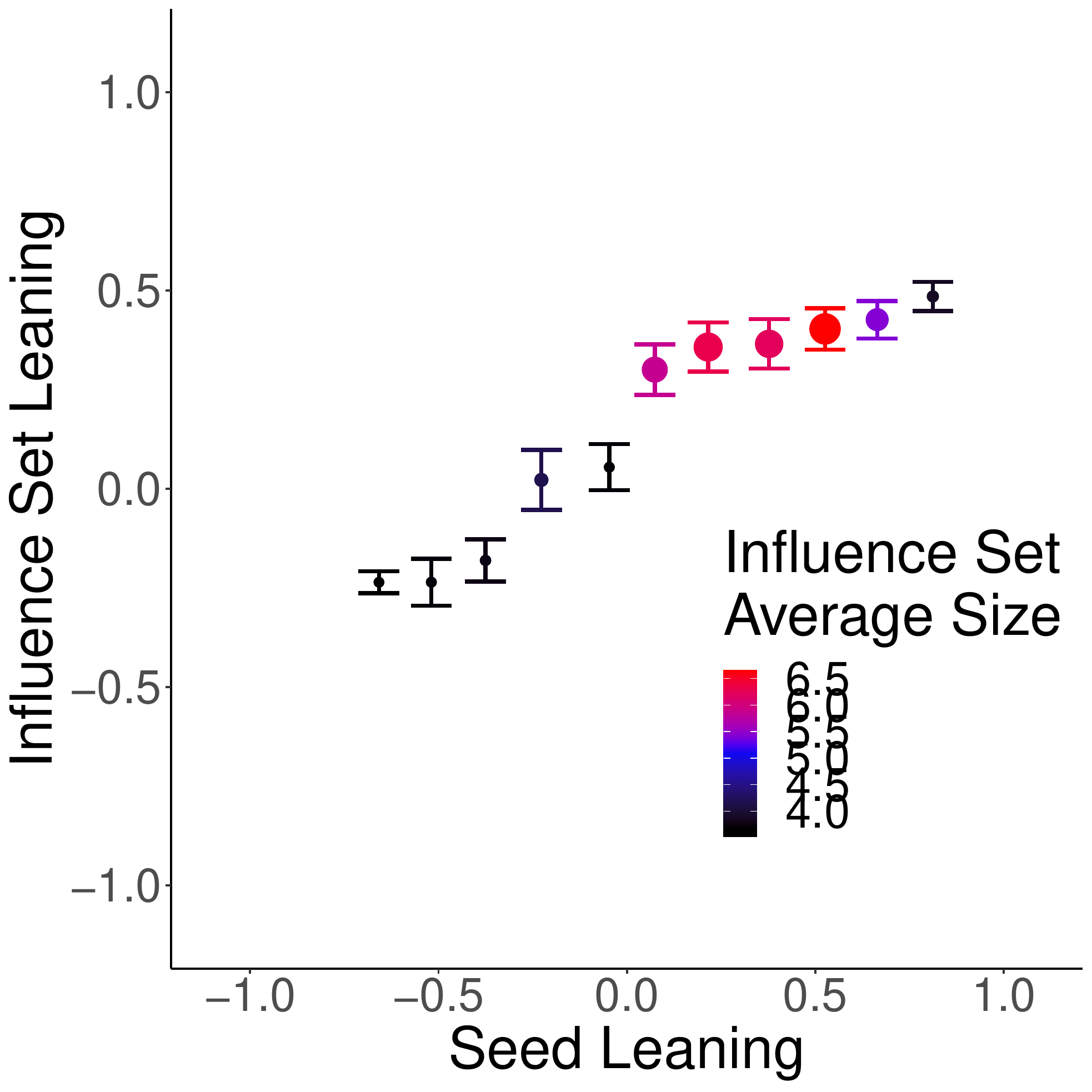}
        \caption{Abortion (Twitter)}
        \end{subfigure}
        \begin{subfigure}[t]{0.3\textwidth}   
            \centering 
            \includegraphics[width=\textwidth]{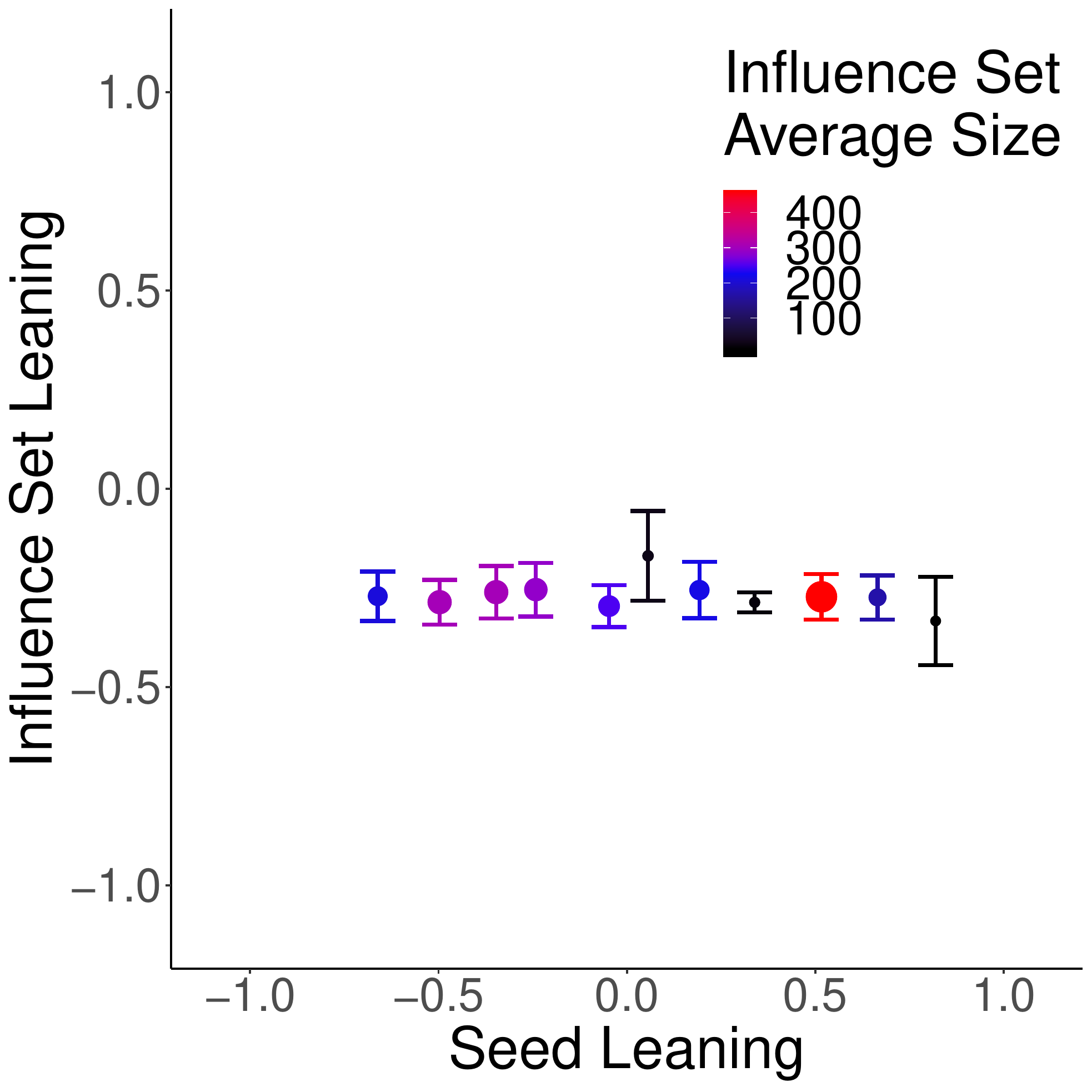}
            \caption{Politics (Reddit)}
        \end{subfigure}
\centering
        \begin{subfigure}[t]{0.3\textwidth}   
            \centering 
            \includegraphics[width=\textwidth]{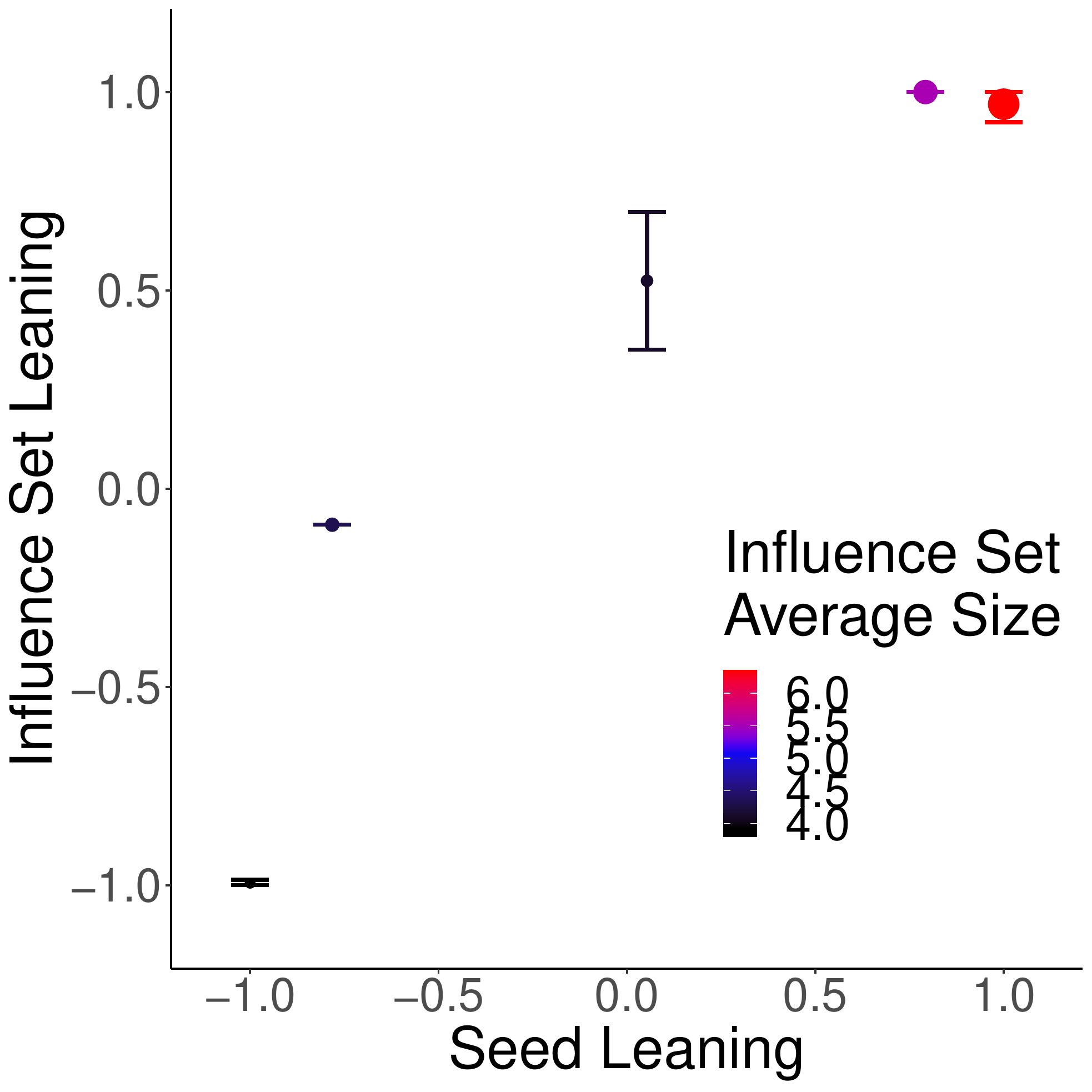}
            \centering
            \caption{Vaccines (Facebook)}
        \end{subfigure}
        \begin{subfigure}[t]{0.3\textwidth}   
            \centering 
            \includegraphics[width=\textwidth]{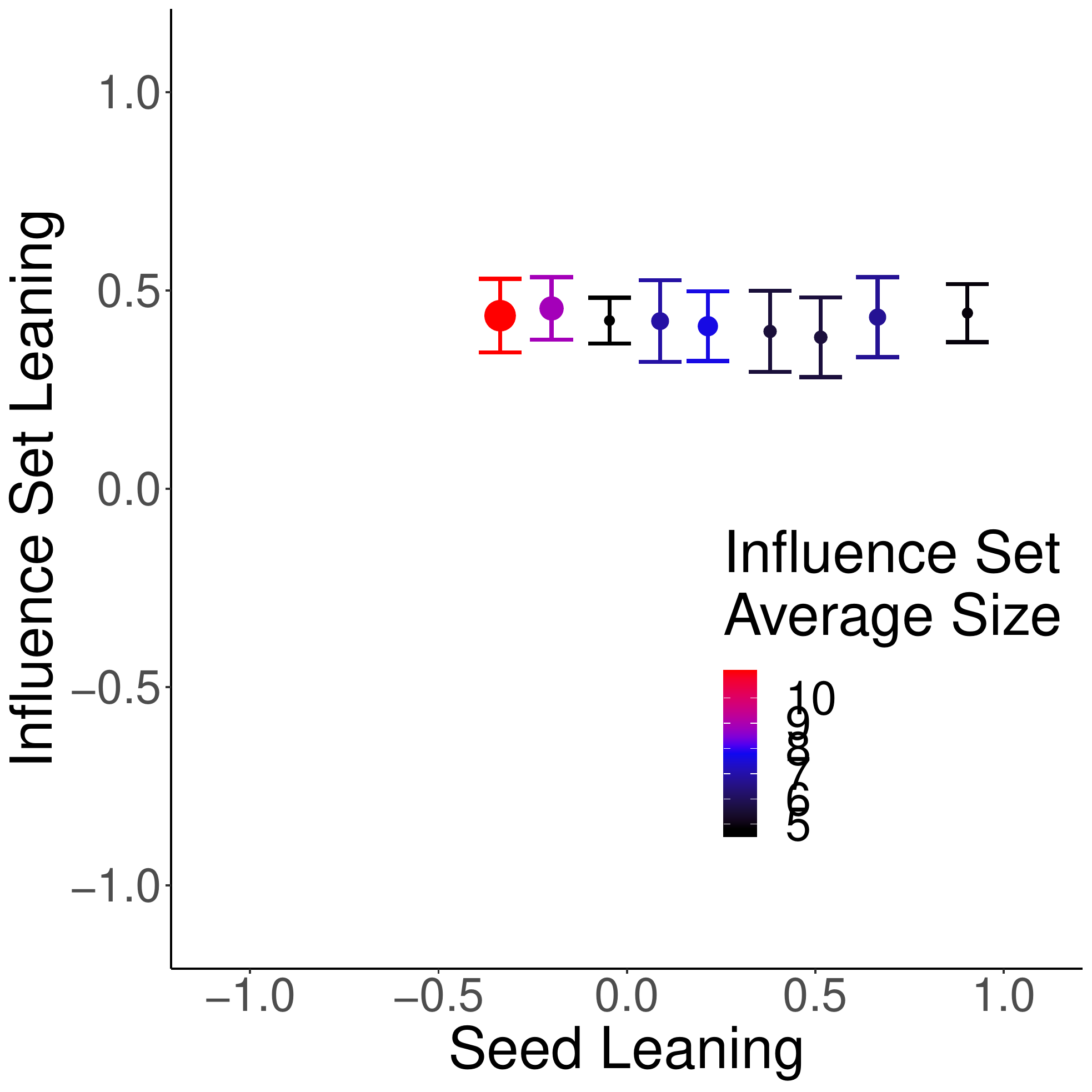}
            \caption{Gab}
        \end{subfigure}
        \centering
        \begin{subfigure}[t]{0.3\textwidth}   
            \centering 
            \includegraphics[width=\textwidth]{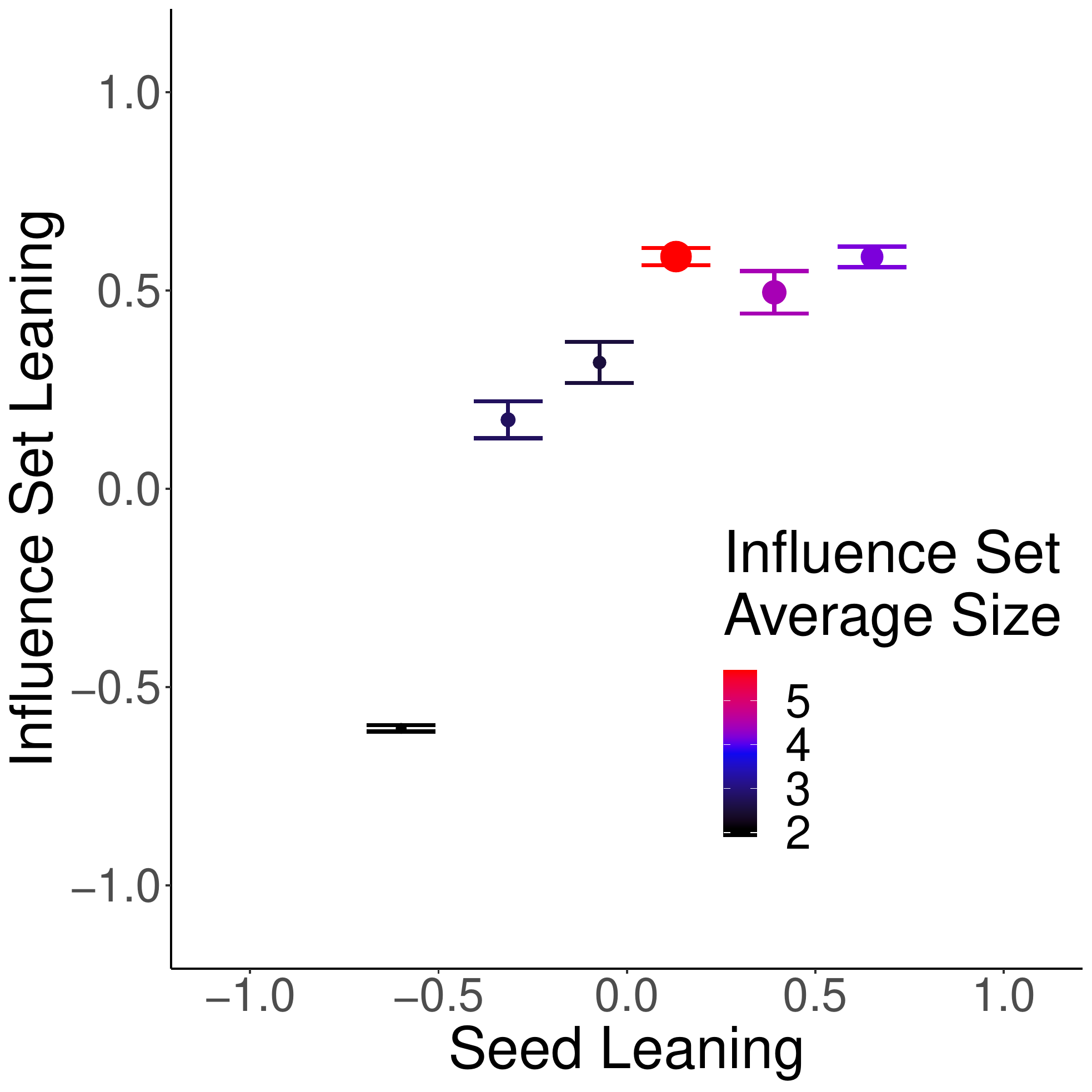}
            \caption{News (Facebook)}
        \end{subfigure}
        \begin{subfigure}[t]{0.3\textwidth}   
            \centering 
            \includegraphics[width=\textwidth]{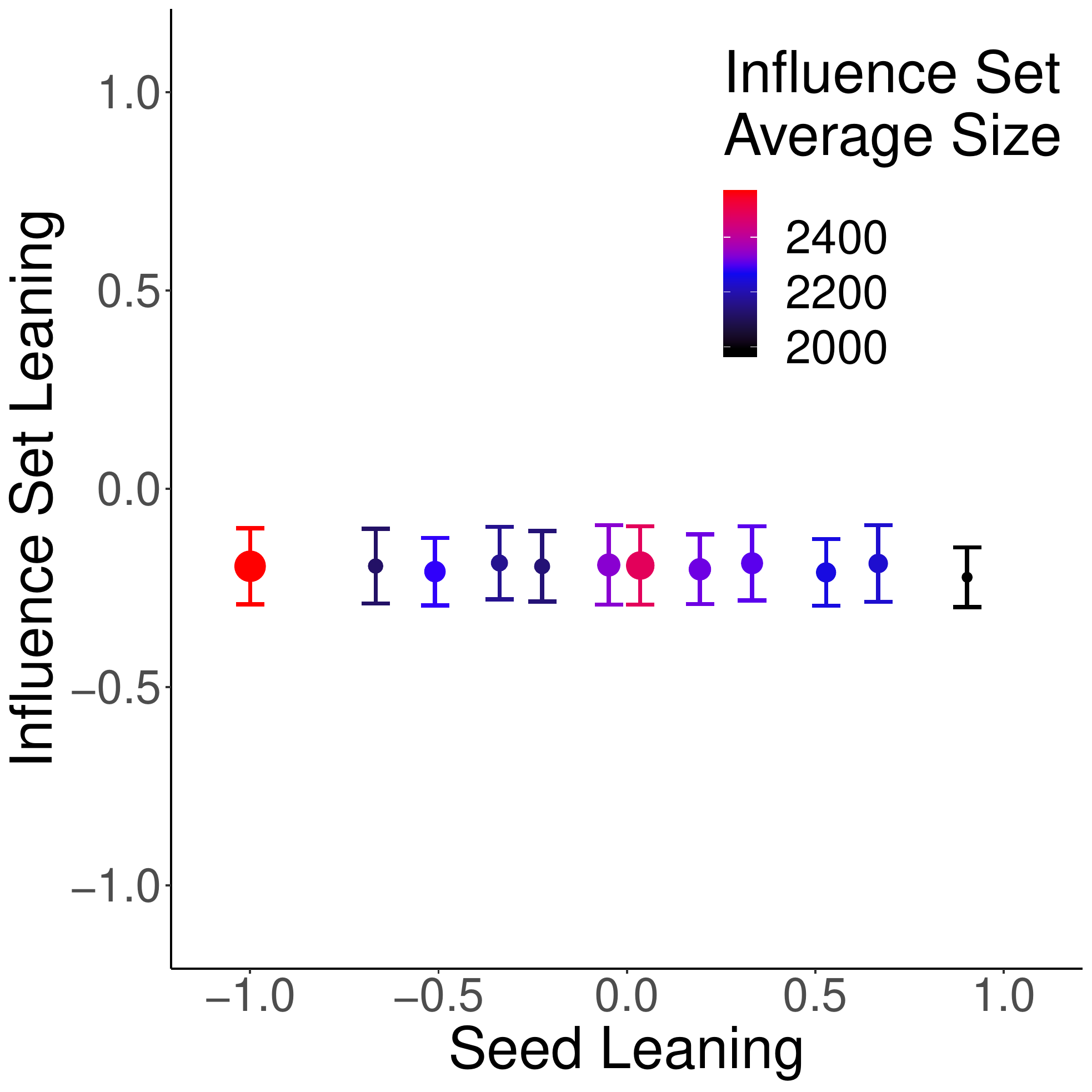}
            \caption{News (Reddit)}
        \end{subfigure}
        \caption{Additional results of the SIR dynamics for the six data sets considered in the main paper. 
        Average leaning $\langle \mu(x) \rangle$ of the influence sets reached by users with leaning $x$, for infection probability 
        $\beta=0.05 \langle k \rangle^{-1}$ (Abortion on Twitter, panel (a)),
        $\beta=0.005 \langle k \rangle^{-1}$ (Politics on Reddit, panel (b)),
        $\beta=0.02 \langle k \rangle^{-1}$ (Vaccines on Facebook, panel (c)),
        $\beta=0.025 \langle k \rangle^{-1}$ (Gab, panel (d)),
        $\beta=0.025 \langle k \rangle^{-1}$ (News on Facebook, panel (e)),
        $\beta=0.01 \langle k \rangle^{-1}$ (News on Reddit, panel (f)),
        while the recovery rate is fixed $\nu = 0.2$.
Size and color of each point is related to the average size of the influence sets. }
        \label{main_four}
\end{figure*}

The results, reported in fig. \ref{main_four}, are qualitatively identical to the ones in the main paper and are reported here for the sake of brevity. Details about the parameters used in the simulations are provided in the caption of Fig. \ref{main_four}.

